\documentclass[a4paper]{article}
\usepackage[utf8]{inputenc}
\usepackage[a4paper, total={16cm, 22cm}]{geometry}

\usepackage{amsmath,amsthm,amssymb,amsfonts}

\usepackage{stmaryrd}

\usepackage{hyperref}

\usepackage{placeins}

\numberwithin{equation}{section}
\usepackage[affil-it]{authblk}
\DeclareMathOperator\arctanh{arctanh}

\usepackage[absolute,overlay]{textpos}
\usepackage{graphicx}
\usepackage{subcaption}

\newcommand{\de}{\,\mathrm{d}}

\newcommand\restr[2]{{
  \left.\kern-\nulldelimiterspace 
  #1 
  \vphantom{\big|} 
  \right|_{#2} 
  }}

\theoremstyle{plain}
\newtheorem{theorem}{Theorem}[section]

\theoremstyle{definition}

\newtheorem{remark}[theorem]{Remark}
\newenvironment{rem}{\begin{remark} \rm}{\end{remark}}

\renewcommand{\sfdefault}{bch}
\def\barr{\hbox{{\fontfamily{\sfdefault}\selectfont I\hskip -.35ex R}}}

\title{Evolution of interface singularities in shallow water equations with variable bottom topography}
\author{R. Camassa${}^1$, R. D'Onofrio${}^2$, G. Falqui${}^{2,4}$, G. Ortenzi${}^{2,4}$, M. Pedroni${}^{3,4}$}

\affil{
{\small ${}^1$University of North Carolina at Chapel Hill, Carolina Center for Interdisciplinary}\\ {\small Applied Mathematics,
Department of Mathematics, Chapel Hill, NC 27599, USA }\\
{\small camassa@amath.unc.edu}
\medskip\\
{\small  $^2$Dipartimento di Matematica e Applicazioni, 
Universit\`a di  Milano-Bicocca, \\ Via Roberto Cozzi 55, I-20125 Milano, 
Italy
}\\
{\small  r.donofrio1@campus.unimib.it, gregorio.falqui@unimib.it, giovanni.ortenzi@unimib.it}
\medskip\\
{\small $^3$Dipartimento di Ingegneria Gestionale, dell'Informazione e della Produzione, 
\\Universit\`a di Bergamo, Viale Marconi 5, I-24044 Dalmine (BG), 
Italy}\\
{\small marco.pedroni@unibg.it}
\medskip\\
{\small  $^4$ INFN, Sezione di Milano-Bicocca, Piazza della Scienza 3, 20126 Milano, Italy}
}

\begin{document}

\maketitle

\begin{abstract}
Wave front propagation with non-trivial bottom topography is studied within the formalism of hyperbolic long wave models. Evolution of non-smooth initial data is examined, and in particular the splitting of singular points and their short time behaviour is described. In the opposite limit of longer times, the local analysis of wavefronts is used to estimate the gradient catastrophe formation and how this is influenced by the topography. The limiting cases when the free surface intersects the bottom boundary, belonging to the so-called ``physical" and ``non-physical" vacuum classes, are examined. Solutions expressed by power series in the spatial variable lead to a hierarchy of ordinary differential equations 
for the time-dependent series coefficients, which are shown to reveal basic  differences between the two vacuum cases: for non-physical vacuums, the equations of the hierarchy are recursive and linear past the first two pairs, while for physical vacuums the hierarchy is non-recursive, fully coupled and nonlinear. The former case may admit solutions that are free of singularities for nonzero time intervals, while the latter is shown to develop non-standard velocity shocks instantaneously. Polynomial bottom topographies simplify the hierarchy, as they contribute only a finite number of inhomogeneous forcing terms to the equations in the recursion relations. However, we show that truncation to finite dimensional systems and polynomial solutions is in general only possible for the case of a quadratic bottom profile. In this case the system's evolution can reduce to, and is completely described by, a low dimensional dynamical system for the time-dependent coefficients.  This system
encapsulates all the nonlinear properties of the solution for general power series initial data, and in particular governs the loss of regularity in finite times at the dry point. For the special case of parabolic bottom topographies,
an exact, self-similar solution class is introduced and studied to illustrate via closed form expressions the general results. 
%

\end{abstract}

\section{Introduction}
Bottom topography, and, in general, sloped boundaries add a layer of difficulty to the study of the hydrodynamics of water waves, and as such have been a classical subject of the literature, stemming from the seminal papers by Carrier and Greenspan~\cite{Carrier58,Greenspan58}, and Gurtin~\cite{Gurtin75}, among others. In fact, the mathematical modelling of liquid/solid moving contact lines is fraught with difficulties from a continuum mechanics viewpoint, and simplifying assumptions are invariably needed when such setups are considered. Nonetheless, even with the simplest models such as the hyperbolic shallow water equations (SWE), 
 fundamental effects such as shoaling of the water layer interacting with bottom topography can be qualitatively, and sometimes quantitatively, described~\cite{Carrier58,Greenspan58}.
Our work follows in these footsteps, and focusses on the dynamics of singular points in the initial value problem for hydrodynamic systems, including that of a contact line viewed as a ``vacuum" point, according to the terminology of gas-dynamics often adopted for hyperbolic systems (see, e.g.,~\cite{Liu1996,Liu1980,CoutardShkoller2011}).

Specifically, unlike most of the existing literature, we study initial data that are not necessarily set up as moving fronts; the evolution of higher order initial discontinuities in these settings can give rise to splitting of singular points, which in turn can lead to gradient catastrophes in finite times. We provide asymptotic time estimates from local analysis as well as characterize the initial time evolution of splitting singular points. Further, we introduce and analyze a class of exact self-similar solutions that illustrate, within closed-form expressions, the generic behaviour mentioned above. In particular,  sloshing solutions for water layers in parabolic bottoms are shown to exist at all times depending on the value of the initial free surface curvature with respect to that of the bottom, while global, full interval gradient catastrophes can occur when the relative values are in the opposite relation. The analysis of these solutions builds upon our previous work in this area~\cite{Camassa-dambreak,Camassa-wetting-mechanism,Camassa-Geom}, extending it to non-flat horizontal bottoms, and highlights the effects of curvature in the interactions with topography. Remarkably, while this solution class is special, it can in fact govern the evolution of more general, analytic  initial data setups, at least  up to a gradient catastrophe time~\cite{Camassa-wetting-mechanism}.

This paper is organized as follows.  We introduce basic elements and notations for the theory of wave fronts in Section~\ref{section: wavefront analysis}, where we consider the 
evolution out of non-smooth initial data. Vacuum (dry) points evolution is studied  in Section~\ref{section:vacuum points}. The dynamics of self-similar solutions interacting 
with a class of bottom topographies is introduced in Section~\ref{section: parabolic solutions}, and a brief discussion and conclusions are presented in Section~\ref{conc}.  The appendices report technical details and results on limiting behaviors. Specifically,  
Appendix~A reviews the physical foundation of the SWE with variable bottom,  Appendix~B reviews the salient points of the adapted variable 
method used in the paper's body, and Appendix~C presents some new, though somewhat technical, results on the parabolic fronts with flat 
bottom topography. 
Finally, online supplementary material complements the exposition by showing animations of representatives of the exact sloshing solutions we derived.


\section{Wavefront analysis}
\label{section: wavefront analysis}
\begin{figure}[t]
\centering
\includegraphics[width=0.5\textwidth]{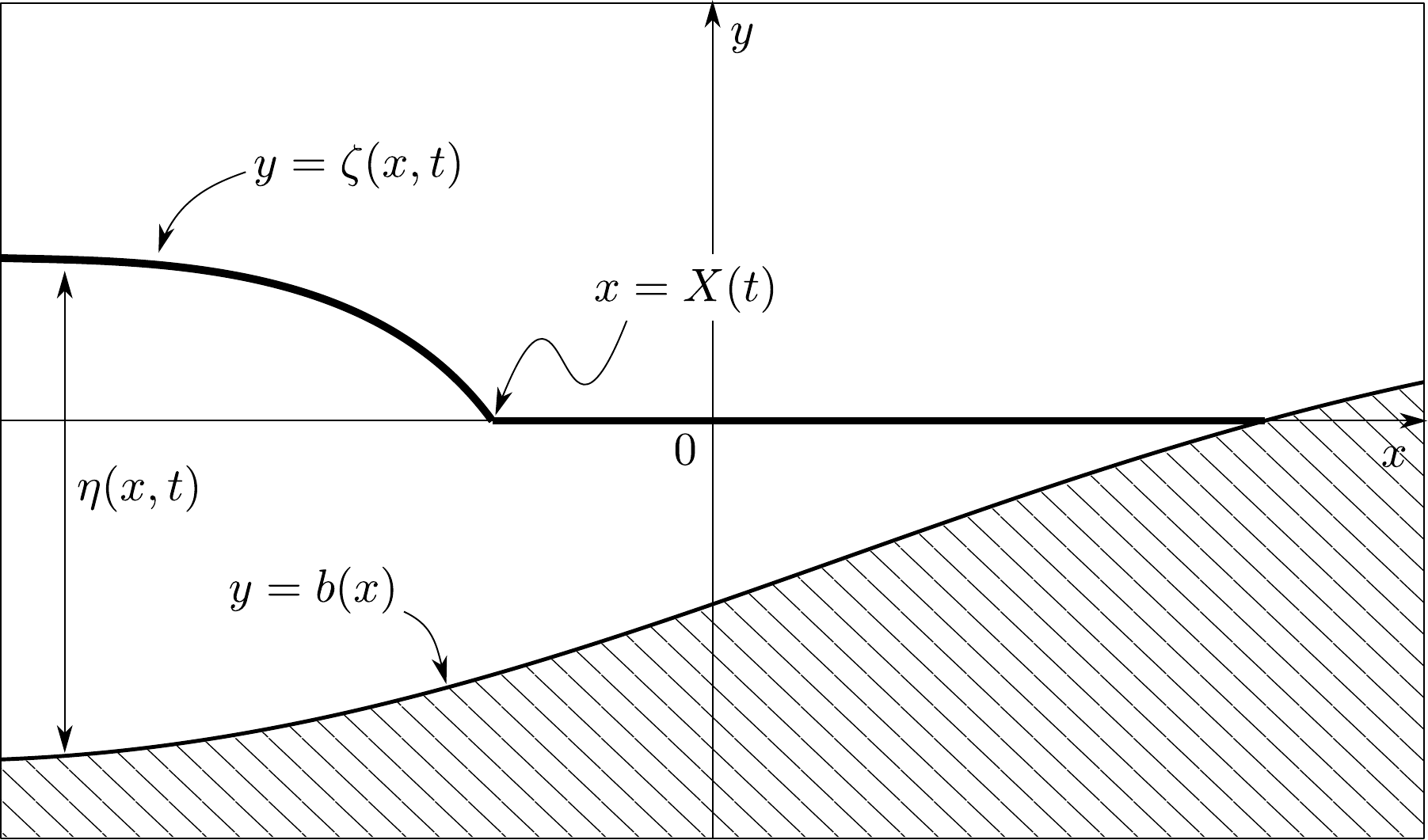}
\caption{A wavefront $x=X(t)$ moving to the right, i.e., $\dot X>0$. It bounds a constant region on its right, where the fluid is quiescent, and a wave region on its left.}
\label{fig:wavefront still fluid}
\end{figure}

The shallow water equations (SWE) studied in this paper are
\begin{equation}
    \label{SWE}
    \eta_t+(u\eta)_x=0,\qquad u_t+u u_x+(\eta+b)_x=0,
\end{equation}
where $u(x,t)$ is the horizontal mean speed, $\eta(x,t)$~is the layer thickness and~$b(x)$ is a smooth function describing the bottom topography. 
Sometimes it is preferable to use the free surface elevation~$\zeta \equiv \eta+b$ instead of the 
layer-thickness~$\eta$ (see figure~\ref{fig:wavefront still fluid}) so that system~(\ref{SWE}) becomes
\begin{equation}
    \label{SWE-z}
    \zeta_t+(u(\zeta-b))_x=0,\qquad u_t+u u_x+\zeta_x=0\, .
\end{equation}

This section is devoted to the so-called wavefront expansion method, which is a helpful tool for analyzing the breaking of hyperbolic waves, especially when the system at hand is non-autonomous/non-homogeneous.
The idea is to study the solution in a neighborhood of the front location $x=X(t)$ by the shifting map (see e.g., \cite{Whitham99,Moodie91,Greenspan58,Gurtin75} and Appendix B for a review)
\begin{equation}
    \xi=x-X(t),\qquad \tau=t.
\label{adap-var}
\end{equation}
In the configuration depicted in Figure~\ref{fig:wavefront still fluid}, on the left of the moving front location~$x=X(t)$ a generic function~$f(\xi,\tau)$ can be assumed to be expanded in Taylor series
\begin{equation}
\label{sol-exp}
    f=f_0+f_1(\tau)\xi+f_2(\tau)\xi^2+\dots,\qquad f_k(\tau)=\lim_{\xi\to 0^-}\frac{1}{k!}\frac{\partial^k f}{\partial \xi^k}(\xi,\tau).
\end{equation}
Equations~(\ref{SWE-z}) give rise to the hierarchy of ODEs
\begin{equation}
\begin{split}
    &\dot \zeta_n+(n+1)\left[(u_0-\dot X)\zeta_{n+1}-b_0 u_{n+1}\right]-(n+1)b_{n+1}u_0+\\
            &\hspace{5cm}+(n+1)\sum_{k=1}^n(\zeta_k-b_k)u_{n+1-k}=0, \\
 & \dot u_n+(n+1)\left[\zeta_{n+1}+(u_0-\dot X)u_{n+1}\right]+\sum_{k=1}^n k \, u_k u_{n+1-k}=0
    \end{split} 
    \label{SWE-ODE}
\end{equation}
where $u_i(t), \zeta(t)$ and $b_i(X(t))$ are the time-dependent coefficients of the $i$-th order term in the $\xi$  expansion~(\ref{sol-exp}).
We focus on the wavefront analysis to extract information about the time evolution of a class of piecewise-defined initial conditions for the SWE over a general bottom. The section is closed with a reminder of the flat bottom case (see also \cite{Camassa-dambreak,Camassa-wetting-mechanism}). In this particular setting, tools like simple waves and Riemann invariants allow for a closed form expression of the complete solution, and the shock time is computed without resorting to an asymptotic wavefront expansion.
%


\subsection{Piecewise smooth initial conditions}
\label{sub: piecewise initial conditions}

In many physical interesting situations, the initial conditions for the SWE are piecewise smooth and globally continuous. The wavefront expansion technique provides a valuable tool for their analysis, especially in the presence of nontrivial bottom topographies. When $b_x=0$, the shallow water system admits Riemann invariants and simple waves. This allows to look for singularities from a global perspective, and compute the first shock time associated with given initial conditions (see \cite{Camassa-wetting-mechanism}). On the other hand, when $b_x\ne 0$, the SWE are inhomogeneous, and Riemann variables are no longer invariant along characteristic curves. Simple waves are unavailable as well, hence for general bottom shapes only the local viewpoint of the wavefront expansion for the study of singularities is in general available for analytical advances. We describe here how the machinery of the wavefront expansion applies to a prototypical example, and  discuss the special case of a flat bottom in the next section.

\begin{figure}
\centering
\includegraphics[width=0.6\textwidth]{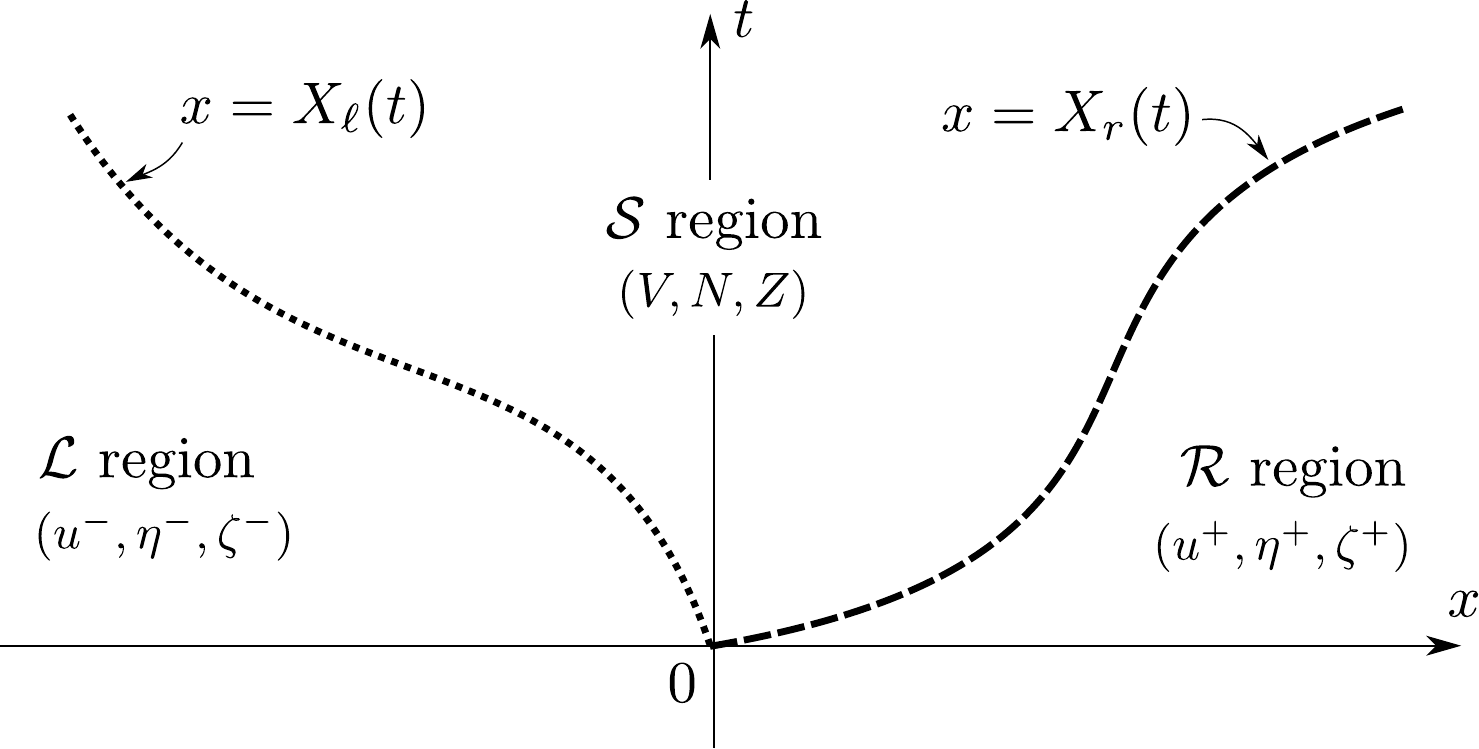}
\caption{Sketch of trajectories of the two wavefronts $x=X_r(t)$ and $x=X_\ell(t)$ arising from piecewise differentiable initial conditions~(\ref{pw initial conditions u})--(\ref{pw initial conditions y}). This data class identifies three distinct regions, $\cal{L}$, $\cal{S}$ and $\cal{R}$, in the spacetime plane. The field variables of the SWE are denoted differently in each of the three regions so defined.}
\label{fig:spacetime partition}
\end{figure}

\begin{figure}
\centering
\includegraphics[width=0.45\textwidth]{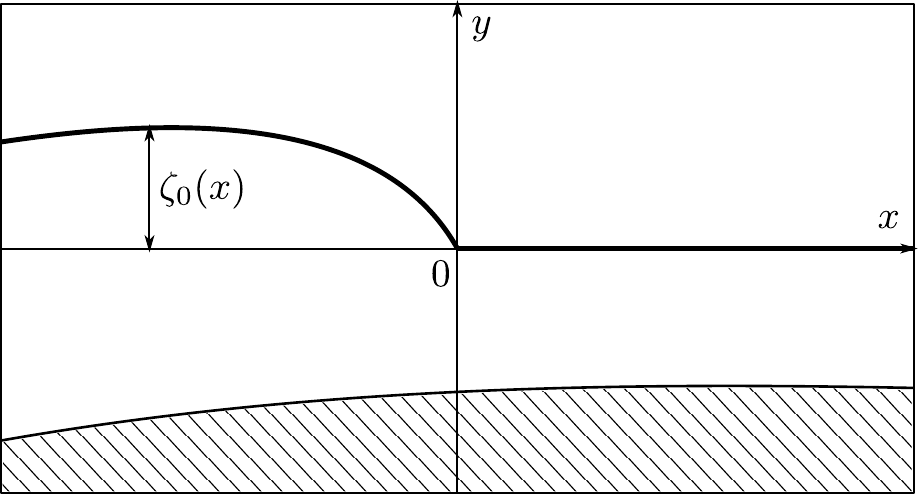}
\includegraphics[width=0.45\textwidth]{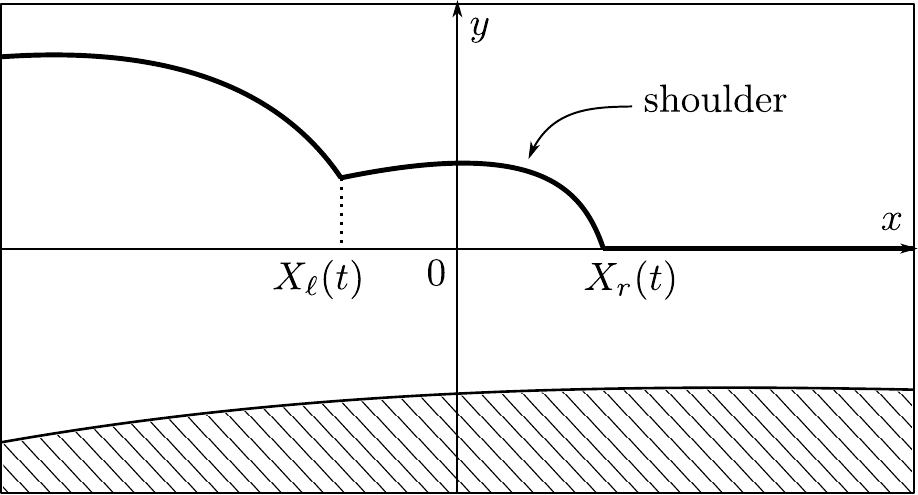}
\caption{Sketch of initial conditions of class~(\ref{pw initial conditions y}) (left) and a snapshot of their evolution at an instant $t>0$ prior to any shock development (right). The shoulder part of the solution, not present at the initial time, is enclosed by the two moving points $x=X_r(t)$ and $x=X_\ell(t)$ for $t>0$.}
\label{fig:shoulder}
\end{figure}

We assign initial conditions as follows: the fluid is initially at rest, and the water surface is composed by a constant part on the right, glued at the origin to a general smooth part on the left, 
\begin{equation}
\label{pw initial conditions u}
    u(x,0)=0\qquad \textnormal{for any }x\in \barr
\end{equation}
and
\begin{equation}
\label{pw initial conditions y}
    \zeta(x,0)=\begin{cases}\zeta_{\mathrm{in}}(x) &\textnormal{for }x\le 0,\\ 0 &\textnormal{for }x>0.\end{cases}
\end{equation}
Here, $\zeta_{\mathrm{in}}(x)$ is a smooth function in $(-\infty,0]$, subject to the condition $\zeta_{\mathrm{in}}(0)=0$, which ensures the global continuity of the initial data. 
We further require that $\zeta_{\mathrm{in}}'(0)\ne 0$ (and bounded), so that the water surface has a discontinuous first derivative, a corner,  at $x=0$ for $t=0$. Finally, we assume that $
\zeta_{\mathrm{in}}(x)> b(x)$ in some neighbourhood of $x=0$. This assumption ensures that the SWE are locally hyperbolic so that a couple of distinct characteristics pass 
through the origin in the spacetime plane (see Figures \ref{fig:spacetime partition} and \ref{fig:shoulder}). These characteristics  satisfy
\begin{equation}
\label{X_r X_ell}
    \begin{cases}\dot X_r=u+\sqrt{\eta}\\ X_r(0)=0\end{cases},\qquad \begin{cases}\dot X_\ell=u-\sqrt{\eta}\\ X_\ell(0)=0\end{cases}.
\end{equation}
As remarked above, derivative jumps propagate along  characteristic curves of  hyperbolic systems 
and the singular point initially located at $x=0$ splits into a pair of singular points transported along $x=X_r(t)$ and $x=X_\ell(t)$ after the initial time. Thus, for $t>0$, and as long as shocks do not arise, the solution is composed by three distinct smooth parts defined by the above pair of characteristics  (see Figure \ref{fig:spacetime partition}). We append the superscript ``$+$'' to variables in the region~$\cal{R}$, that is for $x>X_r(t)$, characterized by the constant state $u^+(x,t)=0$, $y^+(x,t)=0$. Similarly we denote with a ``$-$'' the variables in the region~$\cal{L}$, for $x<X_\ell(t)$, which we regard as known functions of $(x,t)$, i.e., $u^-(x,t),\eta^-(x,t),\zeta^-(x,t)$ are solutions of the SWE corresponding to the initial condition $\zeta(x,0)=\zeta_{\mathrm{in}}(x)$. This construction is well defined in the $\cal{L}$ region, as this is defined  along the negative semiaxis $x<0$. Finally, we denote with capital letters~$V,N,Z$ the solution in the middle region $\cal{S}$, that is for $X_\ell(t)<x<X_r(t)$. We call this portion of the spacetime plane, enclosed by the two singular points, the \textit{shoulder} \cite{Camassa-dambreak,Camassa-wetting-mechanism}.

Even if no singularity were to occur in the $\cal{L}$ region, loss of regularity could certainly happen in the shoulder region~$\cal{S}$. Although we cannot generally rule out the onset of singularities in the
 interior of $\cal S$, we focus here on its boundaries, i.e. the characteristic curves $x=X_\ell(t)$ and $x=X_r(t)$. The machinery described in the previous paragraph is 
 immediately applicable to the right wavefront $x=X_r(t)$, whereas it requires some further adaptation to be used for $x=X_\ell(t)$, as it propagates across a nonconstant
  state. Thus, we next focus  on the solution near the right boundary of the shoulder region, $x=X_r(t)$.

Piecewise initial conditions~(\ref{pw initial conditions y}) and~(\ref{pw initial conditions u}) introduce an element of novelty with respect to the analysis of Gurtin~\cite{Gurtin75}
mentioned in the previous section. The setting considered by Gurtin consists of a travelling wavefront stemming from some unspecified initial conditions. The wavefront analysis 
 is then applied to this dynamical setting, starting from some generic time, arbitrarily picked as the initial one. On the other hand, if the fluid is initially at rest, the wavefronts 
 arise as a consequence of the fluid relaxation. The shoulder components of the solution are not present at the initial time, so special attention is needed to identify initial conditions for the wavefront equations defined 
by the first order of system~(\ref{SWE-ODE}) (see~(\ref{Riccati u_1 y_1}) for details)
 \begin{equation}
 \dot u_1+\frac{3}{2}u_1^2+\frac{5\ddot X}{2\dot X}u_1=0,\qquad \dot \zeta_1+\frac{3}{2\dot X}\zeta_1^2+\frac{3\ddot X}{2\dot X}\zeta_1=0.
\end{equation}
Note that neither $u_1$ or $\eta_1$ are defined at $t=0$, as they represent the limits of the 
 relevant quantities for $x\to X_r(t)^-$ with $X_\ell(t)<x<X_r(t)$. However, a consistent definition for $u_1(0)$ and $\zeta_1(0)$ can still be given in terms of appropriate limits 
 as $(x,t)\to (0,0)$ from the interior of $\cal S$, provided the gradients $\nabla Z$ and $\nabla V$ are continuous up to the boundary $\partial \cal S$. Namely we define
\begin{equation}
    \begin{cases}u_1(0)\equiv \lim_{(x,t)\to (0,0)}V_x(x,t)\\ \zeta_1(0)\equiv \lim_{(x,t)\to (0,0)}Y_x(x,t)\end{cases}\quad \textnormal{with } (x,t)\in \cal S.
\end{equation}
These values are computed from the given initial conditions as follows. From the continuity hypothesis on the solution, the fields $Z,V$, as well as their gradients, can be extended to the boundary $\partial \cal S$  along the two characteristics $X_r$, $X_\ell$ by taking the appropriate limits (the same applies to $\zeta^-,u^-,\zeta^+,u^+$ on their respective domains), and we have
%
%
\begin{equation}
    \zeta^-(X_\ell(t),t)=Z(X_\ell(t),t),\qquad 0=Z(X_r(t),t).
\end{equation}
Once differentiated with respect to time these two relations give
\begin{gather*}
    \zeta^-_x(X_\ell(t),t)\dot X_\ell(t)+\zeta^-_t(X_\ell(t),t)=Z_x(X_\ell(t),t)\dot X_\ell+Z_t(X_\ell(t),t),\\
    0=Z_x(X_r(t),t)\dot X_r+Z_t(X_r(t),t).
\end{gather*}
We now subtract the second equation from the first one, and let $t\to 0^+$. The continuity of $Z,Z_x,Z_t$ on $\overline{\cal S}$ allows us to write
\begin{gather*}
    \lim_{t\to 0^+}Z_x(X_\ell(t),t)=\lim_{t\to 0^+}Z_x(X_r(t),t)\equiv \zeta_1(0), \\
    \lim_{t\to 0^+}Z_t(X_\ell(t),t)=\lim_{t\to 0^+}Z_t(X_r(t),t),
\end{gather*}
which yields an expression for the initial value of $\zeta_1$,
\begin{equation}
    \zeta_1(0)=\frac{\zeta_x^-(0,0)\dot X_\ell(0)+\zeta^-_t(0,0)}{\dot X_\ell(0)-\dot X_r(0)}.
    \label{zet10}
\end{equation}
According to (\ref{pw initial conditions u}), the velocity field $u$ vanishes identically at the initial time, so that  $\dot X_r(0)=-\dot X_\ell(0)$ from (\ref{X_r X_ell}). Furthermore, the continuity equation 
{(\ref{SWE-z})}
implies $\zeta_t(x,0)=0$ for
$x\neq 0$, so~(\ref{zet10}) simplifies to
\begin{equation}
\label{shoulder initial slope}
    \zeta_1(0)=\frac{\zeta_x^-(0,0)}{2}\equiv\frac{\zeta_{\mathrm{in}}'(0^-)}{2},
\end{equation}
where we have used the notation $\zeta_{\mathrm{in}}'(0^-)$ to denote a one-sided derivative. 

\begin{figure}[t]
\centering
\includegraphics[width=0.6\textwidth]{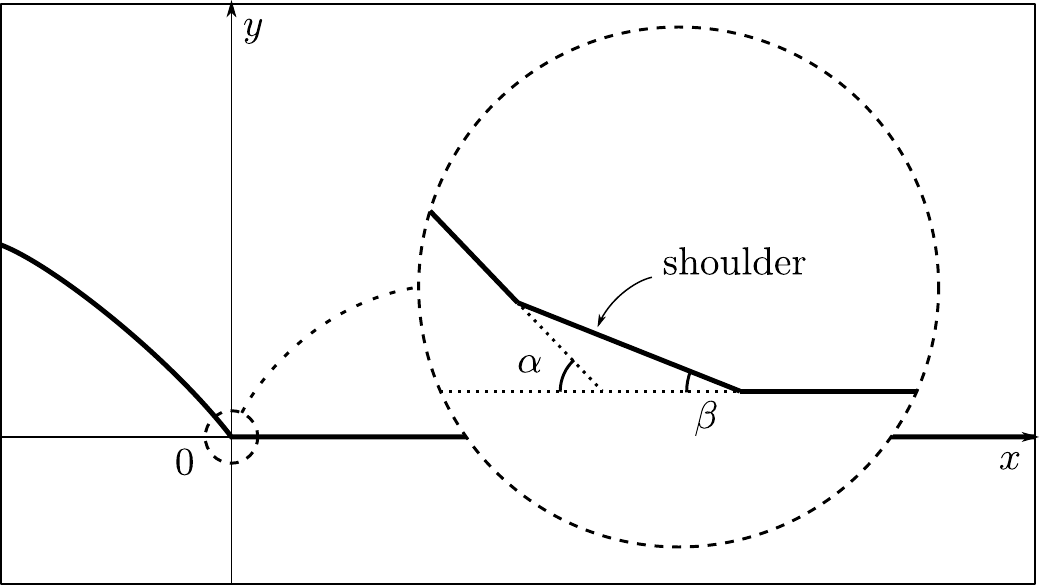}
\caption{Sketch of the solution at a very early stage $t\ll 1$. The shoulder part is approximately a segment with slope $\tan(\beta)\simeq -\zeta_1(0)$,  half of the surface slope in the region $\cal{L}$, $\tan(\alpha)\simeq -\zeta_0'(0^-)$.}
\label{fig:shoulder birth}
\end{figure}

Interestingly, immediately after the initial time, the slope of the water surface in region of the shoulder is half that of the neighboring region $\cal{L}$ (see Figure \ref{fig:shoulder birth}). Thus, according to (\ref{shoulder initial slope}), the initial slope of the shoulder part $\zeta_{\mathrm{in}}'(0^-)$ equals the algebraic mean of the slope values on the two sides.
As $\zeta_1=\dot X_r u_1$, we also obtain the corresponding initial condition for $u_1$,
\begin{equation}
    u_1(0)=\frac{\zeta_{\mathrm{in}}'(0^-)}{2\dot X_r(0)}=\frac{\zeta_{\mathrm{in}}'(0^-)}{2\sqrt{-b(0)}}.
\end{equation}
Here equation~(\ref{wavefront motion}) in Appendix B has been used in the last equality. Further details are reported in this appendix. Thus, together with these initial conditions, (\ref{y_1(t)}) (or (\ref{y_1(x)}))  determine the onset of a gradient catastrophe at the wavefront $x=X_r(t)$. The gradient divergence  depends on the bottom shape $b(x)$ as well as the given initial data $\zeta_{\mathrm{in}}(x)$. Of particular relevance is the sign of the surface slope $\zeta_{\mathrm{in}}'(0^-)$ near the singular point $x=0$, which markedly affects the solution behaviour. We will apply the results of this section to Section~\ref{section: piecewise parabolic solutions}, where a particular form for $\zeta_{\mathrm{in}}(x)$ is prescribed.

\subsection{The case of a flat bottom}
\label{sec: reminder flat bottom}
We now recall ideas already presented in \cite{Camassa-wetting-mechanism} for studying the time evolution of the shoulder region in the particular case $b_x=0$, which identifies a flat, horizontal bottom. It is straightforward to check that under this assumption, the shallow water system becomes autonomous (and homogeneous), which puts many analytical tools at our disposal. Indeed, as the solution features a constant state for $x>X_r(t)$, one of the Riemann invariants is constant throughout the shoulder region where the solution itself is a {simple wave} (see, e.g.,  \cite{Whitham99}). This fact allows us to obtain an exact (implicit) analytical expression for the solution in the shoulder region~$\cal{S}$ and determine when a shock first appears.

\begin{figure}
    \centering
    \includegraphics[width=0.7\textwidth]{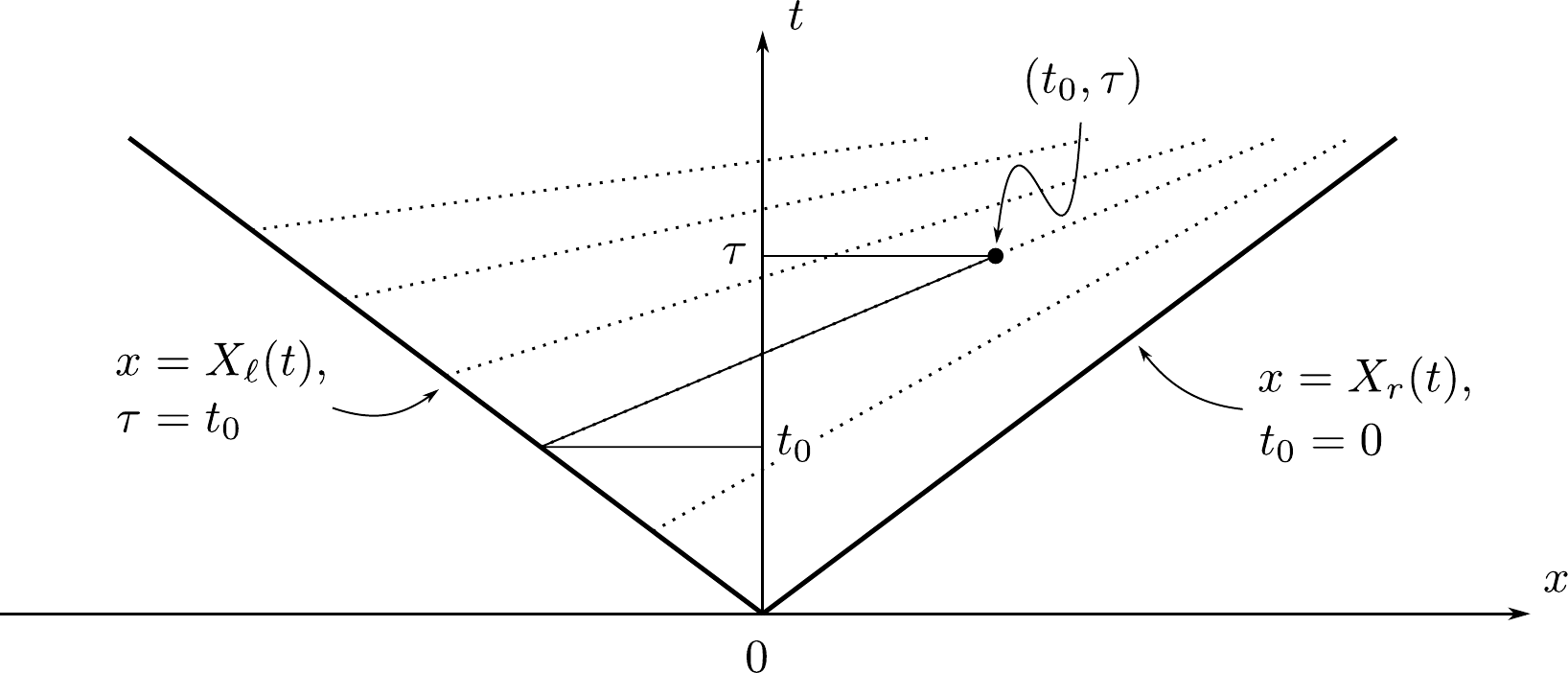}
    \caption{ Space-time diagram of the shoulder region for small times and initial data $u(x,0)\equiv 0$. The right and left limiting characteristics are straight lines of slope $\displaystyle \sqrt{Q}$ 
    and $\displaystyle -\sqrt{Q}$ respective;ly. Dotted lines indicate positive characteristics in the shoulder region, where the construction of the coordinates $(t_0,\tau)$ for a point in the region is depicted; with the  boundaries identified by the conditions $\tau=t_0$ and $t_0=0$.}
    \label{fig: shoulder coordinates}
\end{figure}

For definiteness, we assume that $b(x)=-Q$, with $Q>0$, so the initial conditions (\ref{pw initial conditions y}) in this case read
\begin{equation}
    \eta(x,0)=\begin{cases}
    \eta_{\mathrm{in}}(x) & \textnormal{for }x<0,\\
    Q & \textnormal{for }x>0.
    \end{cases}
\end{equation}
As above, we still assume that the velocity field is identically zero at the initial time. It is worth noting that, for the flat bottom case, the curve $x=X_r(t)$ in Figure \ref{fig:spacetime partition} is a straight line of slope $\displaystyle \sqrt{Q}$, and the curve $x=X_\ell(t)$ has tangent $-\displaystyle\sqrt{Q}$ at its starting point (see Figure \ref{fig: shoulder coordinates}).

By plugging $b_x=0$ into 
{ (\ref{SWE})}, the shallow water system can be cast in the characteristic form
\begin{equation}
\label{characteristic form SWE}
    \partial_t R_\pm+\lambda_\pm\partial_x R_\pm=0,
\end{equation}
where the characteristic velocities $\lambda_\pm$ and the Riemann invariants $R_\pm$ are defined by
\begin{equation}
    \lambda_\pm=u\pm\sqrt{\eta},\qquad R_\pm=u\pm2\sqrt{\eta}.
\end{equation}
As long as shocks do not arise, the Riemann invariant $R_-$ is everywhere constant in the shoulder region. Namely, we have
\begin{equation}
    R_-=V-2\sqrt{N}\equiv -2\sqrt{Q}\quad \textnormal{in}\quad \overline{\cal{S}\cup{R}}.
\end{equation}
On the other hand, $R_+$ is constant along positive characteristics by virtue of (\ref{characteristic form SWE}), so both fields $N$ and $V$ are constant along curves $\dot x=\lambda_+$. This implies that positive characteristics are straight lines in the shoulder region, and their explicit expression is found by integrating the ODE
\begin{equation}
    \frac{dx}{dt}=3\sqrt{N}-2\sqrt{Q}.
\end{equation}
We assign initial conditions for this equation along the left boundary of the shoulder region, that is
\begin{equation}
    x(t_0)=X_\ell(t_0).
\end{equation}
Therefore, the general equation for positive characteristics is
\begin{equation}
\label{characteristics shoulder}
    x=X_\ell(t_0)+(3\sqrt{N}-2\sqrt{Q})(t-t_0).
\end{equation}
Each of the characteristics (\ref{characteristics shoulder}) intersects the left boundary of the shoulder region at a different time $0\le t_0<+\infty$, which serves as a parameter to label the characteristic curves themselves. Note that the right boundary of $\cal S$ is the positive characteristic whose label is $t_0=0$. Moreover, as long as shocks do not arise, the parameter $t_0$ can be used along with the time variable to build a local coordinate system in the spacetime plane (see Figure \ref{fig: shoulder coordinates}). Namely, we define a change of variables $(t_0,\tau)\mapsto (x,t)$ by
\begin{equation}
\label{shoulder coordinates}
    \begin{cases}
    x=X_\ell(t_0)+(3\sqrt{N}-2\sqrt{Q})(\tau-t_0),\\
    t=\tau.
    \end{cases}
\end{equation}
This gives a concise expression of the solution in the shoulder region. Indeed, as already stated, both fields $N$ and $V$ are constant along positive characteristics, so they are independent of $\tau$,
\begin{equation}
    N=N(t_0),\qquad V=V(t_0).
\end{equation}
Furthermore, as long as the solution is continuous, the value of the field variables is fixed by the solution $\eta^-$, $u^-$ in the $\cal{L}$ region. Namely, we can write
\begin{equation}
\label{shoulder solution}
    N(t_0)=\eta^-(X_\ell(t_0),t_0),\qquad V(t_0)=u^-(X_\ell(t_0),t_0).
\end{equation}
Once the solution in the shoulder region is known, we can look for gradient catastrophes as follows. First, we express the space and time derivatives in the new coordinates as
\begin{equation}
    \partial_x={1\over x_{t_0}}\partial_{t_0},\qquad \partial_t=\partial_\tau-\frac{x_\tau}{x_{t_0}}\partial_{t_0},
\end{equation}
where
\begin{equation}
    \begin{cases}
    x_{t_0}(t_0,\tau)=\dot X_\ell(t_0)+3\left(\sqrt{N(t_0)}\right)_{t_0}(\tau-t_0)-3\sqrt{N(t_0)}+2\sqrt{Q}\,,\\
    x_\tau(t_0)=3\sqrt{N(t_0)}-2\sqrt{Q}\,.
    \end{cases}
\end{equation}
Therefore, the slope of the water surface in the shoulder region is
\begin{equation}
    N_x(t_0,\tau)=\frac{N_{t_0}(t_0)}{x_{t_0}(t_0,\tau)},
\end{equation}
which shows that the appearance of a gradient catastrophe is identified by the vanishing of $x_{t_0}$. The condition $x_{t_0}=0$ gives the time when the positive characteristic starting at $t=t_0$ from the left boundary of the shoulder region intersects another characteristic of the same family, that is 
\begin{equation}
\label{shock time single}
    \bar{\tau}(t_0)=t_0+\frac{3\sqrt{N(t_0)}-2\displaystyle \sqrt{Q}-\dot X(t_0)}{3\left(\sqrt{N(t_0)}\right)_{t_0}}.
\end{equation}
The earliest shock time for the shoulder part of the solution equals the non-negative infimum of the shock times~(\ref{shock time single}) associated to the single characteristics  under the additional constraint that the shock position lies within the shoulder region itself, \begin{equation}
\label{shock time inf}
    \tau_\textnormal{sh}=\inf_{t_0\in I}\bar{\tau}(t_0),
\end{equation}
where the set $I$ is 
\begin{equation}
    I\equiv\left\{t_0:X_\ell(\bar{\tau}(t_0))\le x(t_0,\bar{\tau}(t_0))\le X_r(\bar{\tau}(t_0))\right\}.
\end{equation}
The shock time $\tau_\textnormal{sh}$ is computed in the context of a specific example in Appendix~C.

Thanks to the solution's availability in the case of a flat bottom, we can verify the assumptions we made in the previous section for the general case. First of all, it is straightforward to check that the gradient of the field variables is continuous at the intersection of the shoulder domain with a strip 
$\barr \times[0,\epsilon]$ 
with $\epsilon$ sufficiently small. For example, the gradient of $N(x,t)$ (as well as of $V(x,t)$),
\begin{equation}
\label{gradient N}
    N_x=\frac{N_{t_0}(t_0)}{x_{t_0}},\qquad N_t=-\frac{x_\tau}{x_{t_0}}N_{t_0}(t_0),
\end{equation}
is continuous up to the boundary of the shoulder domain as long as shocks do not arise. Indeed, $x_{t_0}$ is well defined on both the boundaries $t_0=\tau$ and $t_0=0$:
\begin{align}
    &\textnormal{along $x=X_\ell(\tau)$:}\qquad x_{t_0}(\tau,\tau)=\dot X_\ell(\tau)-3\sqrt{N(\tau)}+2\sqrt{Q},\\
    &\textnormal{along $x=X_r(\tau)$:}\qquad x_{t_0}(0,\tau)=-2\sqrt{Q}+3\frac{\de\sqrt{N(t_0)}}{\de t_0}\bigg|_{t_0=0}\tau.
\end{align}
Furthermore, as $\dot X_\ell(0)=-\sqrt{Q}$ and $N(0)=Q$, one can see that
\begin{equation}
    x_{t_0}(0,0)=-2\sqrt{Q}.
\end{equation}
Thus $x_{t_0}$ is continuous and different from zero up to the boundaries of the shoulder domain for sufficiently small times, and so are the gradients $\nabla N$ and $\nabla V$.

Secondly, in this example a direct computation can be performed of the water surface's slope in the shoulder region at the initial time. Indeed, by using (\ref{shoulder solution}), we get
\begin{equation}
    N_{t_0}(0)=\frac{\de\eta^-(X_\ell(t_0),t_0)}{\de t_0}\bigg|_{t_0=0}=\eta^-_t(0,0)+\dot X_\ell(0) \eta^-_x(0,0).
\end{equation}
On the other hand, the initial condition $u(x,0)=0$, combined with the continuity equation, implies $\eta^-_t(0,0)=0$, so that we have
\begin{equation}
    N_{t_0}(0)=-\sqrt{Q}\;\eta_{\mathrm{in}}'(0).
\end{equation}
Therefore, the first of equations (\ref{gradient N}) yields
\begin{equation}
    N_x(0,0)=\frac{\eta_{\mathrm{in}}'(0)}{2},
\end{equation}
in accordance with~(\ref{shoulder initial slope}).

\section{Vacuum points}
\label{section:vacuum points}

The problem of predicting the motion of a shoreline in the presence of waves has a long history. When the bottom has a constant slope, so that $b_{xx}=0$, the SWE can be linearized by the so-called Carrier-Greenspan transform \cite{Carrier58, Rybkin20}. Several closed-form solutions can be constructed in this case and the corresponding shoreline motion can be described exactly, offering an elegant global perspective in the study of the breaking of waves. However, this 
hodograph-like transformation is no longer available for general bathymetry. Hence, we turn to wavefront asymptotics to provide information about the shoreline motion and the local behaviour of the solution.
\begin{figure}[t]
\centering
\includegraphics[width=0.5\textwidth]{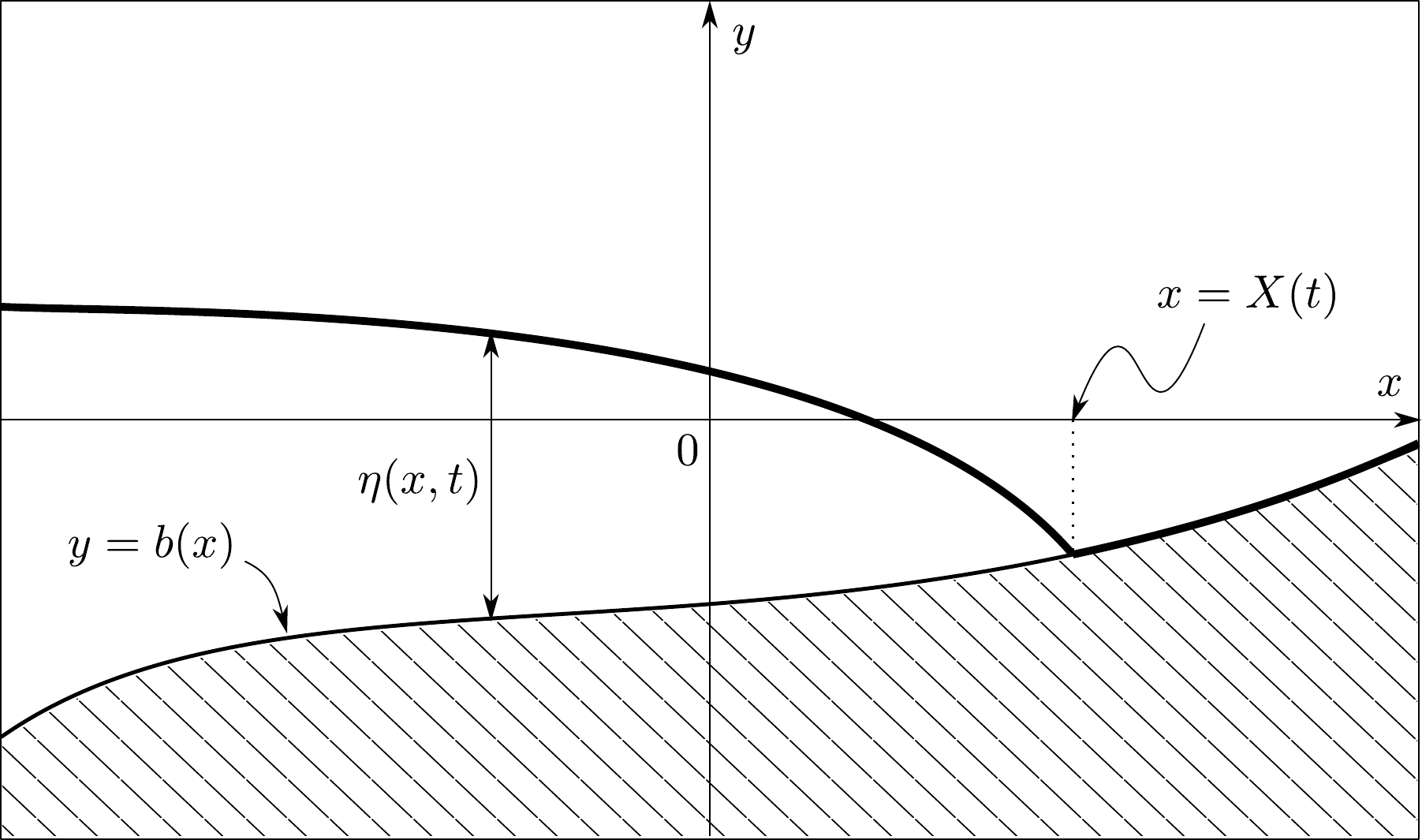}
\caption{Schematic of moving vacuum point $x=X(t)$, separating a fluid-filled region on its left, where the water surface is strictly above the bottom, and a vacuum region to its right, where surface and bottom curves coincide.}
\label{fig:wavefront dry point}
\end{figure}

The shoreline reduces to a single dry point in our one-dimensional setting. In the terminology of gas-dynamics often used for hyperbolic systems, $\eta(x,t)$  plays the role of a gas density and this dry point is referred to as a ``vacuum point," i.e.,  a moving location $x=X(t)$ such that the condition 
$\eta(X(t),t)=0$, holds identically in some interval of time. 
We consider the setup depicted in Figure \ref{fig:wavefront dry point}, where the shoreline $x=X(t)$ separates a fluid-filled region to its left from a vacuum region on its right. For this reason, the term {``vacuum boundary"} will also be used when referring to $x=X(t)$.

{Being a boundary point, it is to be expected that the shoreline's velocity~$\dot{X}(t)$ equals the velocity $u_0$ of the fluid at $X$, i.e., the vacuum boundary is transported by the velocity field.   
(A formal proof can be found,  e.g., in~\cite{Camassa19} for the flat bottom case; however the proof is  independent from the bottom topography $b(x)$, being based only on the local analysis of the mass conservation law for the layer thickness.)}

 In this setting, it is more convenient to use the first form~(\ref{SWE}) of the shallow water equations,
as dry states are simply identified by the condition $\eta=0$. As to the velocity field, there is no preferential or meaningful way of assigning it in the dry/vacuum region. In particular, the zero constant solution is no longer  admissible, since $u$ has to satisfy the non-homogeneous Hopf equation,
\begin{equation}
\label{hopf}
    u_t+u u_x+b_x=0.
\end{equation}
This is the only constraint on $u$ in the dry region, and we will usually leave it unspecified. We still adopt the coordinate system (\ref{adap-var}), so that the SWE read
\begin{equation}
\label{SWE wavefront coordinates}
    \eta_\tau-\dot X\eta_\xi+(u\eta)_\xi=0,\qquad u_\tau-\dot X u_\xi+u u_\xi+\eta_\xi+b_\xi=0. 
\end{equation}
{Next, the wavefront expansion of the solution is introduced for the region~$\xi<0$.  We consider a power series representation for $\eta$, $u$ and $b$ 
in the form (\ref{sol-exp}) (see also  (\ref{solution expansion y}) and (\ref{b expansion}) in Appendix B).
Notice that $\eta_0=0$, in order to agree with the dry-point  condition $\eta(X(\tau),\tau)=0$. In the following, we will abuse notation a little and use $t$ in place of $\tau$ for the time dependence argument, as long as this does not generate confusion. Plugging the previous formal series
in equations (\ref{SWE wavefront coordinates}), and collecting like powers of $\xi$, gives the following infinite hierarchy of ODEs: 
\begin{equation}
\dot{u}_0+b_1+\eta_1=0\,,
\label{hie0}
\end{equation}
 for $n=0$, and 
\begin{gather}
\label{hierarchy eta}
    \dot \eta_n+(n+1)(u_0-\dot X)\eta_{n+1}+(n+1)\sum_{k=1}^n u_k\eta_{n+1-k}=0,\\
    \dot u_n+(n+1)(u_0-\dot X)u_{n+1}+(n+1)(\eta_{n+1}+b_{n+1})+\sum_{k=1}^n k u_l u_{n+1-k}=0\,,
\label{hierarchy u wet}
\end{gather}
for $n>0$. Note that for $n=0$ system~(\ref{SWE wavefront coordinates}) yields a single equation, as 
the $\eta$-equation in~(\ref{SWE wavefront coordinates}) is automatically satisfied and hence provides no information.
As seen above, since~$\dot X=u_0$, the hierarchy simplifies to
\begin{gather} 
\label{hierarchy eta simplified}
    \dot \eta_n+(n+1)\sum_{k=1}^n u_k\eta_{n+1-k}=0 \tag{I},\\
    \dot u_n+(n+1)(\eta_{n+1}+b_{n+1})+\sum_{k=1}^n k\, u_l u_{n+1-k}=0 \tag{II}.
\label{hierarchy u simplified}
\end{gather}
Table~\ref{table:1} shows the explicit form of the first few equations in this hierarchy, corresponding to powers 
of~$\xi$ up to cubic.
\begin{table}
\centering
\small
\begin{tabular}{|l|l|l|}
\hline
 \multicolumn{1}{|c|}{$n$} & \multicolumn{1}{|c|}{(I)} & \multicolumn{1}{|c|}{(II)} \\ 
 \hline
 0 & $0=0$ & $\dot u_0+b_1+\eta_1=0$ \\  
 1 & $\dot \eta_1+2u_1\eta_1=0$ & $\dot u_1+2b_2+u_1^2+2\eta_2=0$ \\
 2 & $\dot \eta_2+3u_2\eta_1+3u_1\eta_2=0$ & $\dot u_2+3b_3+3u_1u_2+3\eta_3=0$ \\
 3 & $\dot \eta_3+4u_3\eta_1+4u_2\eta_2+4u_1\eta_3=0$ & $\dot u_3+4b_4+2u_2^2+4u_1u_3+4\eta_4=0$ \\
 \vdots & \vdots & \vdots\\
 \hline
\end{tabular}
\caption{The first equations~(I)--(II) for the unknown coefficients of the formal series (\ref{sol-exp}) for $u(\xi+X(\tau),\tau)$, $\eta(\xi+X(\tau),\tau)$ and $b(\xi+X(\tau))$. 
}
\label{table:1}
\end{table}

\begin{rem}
\label{rmgen}
A few features of hierarchy~(\ref{hierarchy eta})--(\ref{hierarchy u wet}) are worth a close look:
\begin{enumerate}
\item
 With $u_0=\dot X$ the $\dot u_0$-equation can be written as 
\begin{equation}
\label{wavefront acceleration}
    \ddot X+b_1+\eta_1=0 \,.
\end{equation}
This relates the acceleration of the wavefront (the vacuum boundary) $x=X(t)$ to the slope of the water surface behind it, which is precisely $b_1+\eta_1$.  
\item
A truncation of the infinite hierarchy~(\ref{hierarchy eta simplified})-(\ref{hierarchy u simplified}) to some order $n=N$ would correspond to a reduction of the continuum governed by the SWE to a finite number of degrees of freedom dynamics, with $\eta$ and $u$ being polynomial functions of $x$, respectively of order $N+1$ and $N$. To this end, it is clear that a necessary condition for this to happen is that the bottom topography be a polynomial of degree~$N$, as opposed to an infinite power series. This is reflected by the structure of the hierarchy~(\ref{hierarchy u simplified}), which would lose the terms $\{b_n\}$ (generated by the bottom topography for $n>N$) that make equations~(\ref{hierarchy u simplified}) inhomogeneous. Homogeneity would allow to set the corresponding series coefficients~$\{\eta_{n+1},u_n\}$ to zero if so initially, thereby reducing the power series solutions to mere $\xi$-polynomials.  However, the very same structure of~(\ref{hierarchy u simplified}) shows that the condition of polynomial bottom profiles  in general cannot be sufficient for an exact truncation of the series: even in the absence of the~$\{b_n\}$ terms the equations in hierarchy~(\ref{hierarchy u simplified}) for $n>N$ are not truly homogeneous, since functions of lower index series-coefficient $\{(u_n,\eta_n)\}$ for $n<N$ enter themselves as inhomogeneous forcing functions in all the remaining $n>N$ infinite system.  

\item
The case of a quadratic bottom profile, 
\begin{equation}
b(x)=c_0+c_1 x+c_2x^2/2 
\label{quadb}
\end{equation}
for some constants $c_0$, $c_1$ and $c_2$, say,  is clearly special (and often the one considered in the literature). In fact, for this case $b_1=c_1+c_2\,X(t)$, $b_2=c_2$ (and of course $b_n=0$ for $n>2$). With null 
initial data $\eta_{n+1}(0)$ and $u_n(0)$ for $n>1$, equations~(\ref{hierarchy eta simplified})-(\ref{hierarchy u simplified}) are consistent with $\eta_{n+1}(t)=0$ and $u_n(t)=0$ for $t>0$, $n>1$, and hence the hierarchy truncates to a finite, closed  system for the four unknowns $X(t),\eta_1(t),\eta_2(t)$, and~$u_1(t)$,
\begin{equation}
\label{quadX}
\ddot{X}+c_2 \, X+c_1+\eta_1=0\,,\qquad \dot{\eta_1}+2u_1\eta_1=0\,,\qquad \dot{\eta_2}+3u_1\eta_2=0\,,\qquad
\dot{u_1}+u_1^2+2\eta_2+c_2=0\,.
\end{equation}
This {\it exact} truncation singles out the quadratic case as particularly amenable to complete analysis of solution behaviour, as we shall elaborate in more detail in Section~\ref{section: parabolic solutions} below. 

\item
For $n=1$ equation (\ref{hierarchy eta simplified}) gives
\begin{equation}
    \dot \eta_1+2u_1\eta_1=0 \,;
\end{equation}
this implies that if the initial conditions are such that $\eta_1(0)=0$, then $\eta_1(t)=0$ at all subsequent times, at least for as long as the solution maintains the regularity assumed for the convergence of the power series expressions~(\ref{sol-exp}) 
{for all the variables involved.}
Note that since $\eta(0,t)$ is the layer thickness at the front~$\xi=0$, $\eta_1=0$ implies that the derivative of the free surface matches that of the bottom at the dry point, that is, the free surface is tangent to the bottom there. 
\end{enumerate}
\end{rem}
This last point in Remark~\ref{rmgen} plays an important role in the classification and properties of the solutions with vacuum points, to which we turn next. Specifically, we analyze below the two cases $\eta_1=0$ and $\eta_1<0$ separately; in the literature these two different cases are commonly referred to as the ``nonphysical" and ``physical" vacuum points, respectively~(see, e.g., \cite{Liu1996}), and we will henceforth conform to this terminology.

\subsection{Nonphysical vacuum points}
We first consider the case $\eta_1(0)=0$, for which the water surface is tangent to the bottom at the vacuum boundary.
From equation (\ref{wavefront acceleration}) it follows that the motion of the wavefront is solely determined by the bottom shape as solution to the problem
\begin{equation}
\label{motion of nonphysical vacuum point}
    \ddot X+b_x(X(t))=0,\qquad X(0)=x_0,\qquad \dot X(0)=u_0(0).
\end{equation}
This equation can be integrated once, to get
\begin{equation}
    \frac{\dot X^2}{2}+b(X(t))=\textnormal{constant}.
\end{equation}
Thus, the motion of a nonphysical vacuum point turns out to be the same as that of a (unit mass) particle located at $x$ in a potential $b(x)$: the particle is repelled by local maxima of the bottom topography, and is attracted by local minima. This motion enters the equations of hierarchy~(\ref{hierarchy u simplified}) by providing the $\{b_n(t)\}$ terms, which drive the evolution of the $u$- and $\eta$-coefficients by entering their respective equation as prescribed forcing functions of time, since~(\ref{motion of nonphysical vacuum point}) is no longer  coupled to the $\{u_n,\eta_n\}$ equations.

The special case of a quadratic form for bottom profiles is further simplified for non-physical vacuum. In particular, if $c_2=0$, i.e., the bottom is a straight line,  the first equation in system~(\ref{quadX}) reduces to  
$\ddot X+c_1=0$ and the motion of the vacuum point $X(t)$ is that of uniform acceleration.  If the bottom has a symmetric parabolic shape, $c_1=0$ and the same equation becomes
$\ddot X+c_2 X=0$.
Thus, the motion of the non-physical vacuum point $X(t)$ depends on the concavity of the parabolic bottoms: if it is upward, the motion is oscillatory harmonic, while if it is downward the point $X(t)$ is exponentially repelled from the origin at~$x=0$.

For general bottom profiles that can be expressed as (convergent) power series, the case of non-physical vacuum yields a  remarkable property for the infinite hierarchy~(\ref{hierarchy eta simplified})-(\ref{hierarchy u simplified}). 
When~$\eta_1=0$, the coupling term $u_2\eta_1$ in equation~$(2,\textnormal{I})$ of Table \ref{table:1} is suppressed 
and equations~$(1,\textnormal{II})$ and $(2,\textnormal{I})$ of Table~\ref{table:1}, become a {\it closed} system of two nonlinear differential equations for the two unknowns $u_1(t)$ and $\eta_2(t)$,
\begin{equation}
\label{u1 eta2 system}
    \dot u_1+u_1^2+2\eta_2+2b_2=0,\qquad \dot \eta_2+3u_1\eta_2=0.
\end{equation}
As remarked above, the function $b_2(t)$ can be thought of as an assigned time dependent forcing function,  defined by
\begin{equation}
\label{b2t}
    b_2(t)=\frac{1}{2}\,b_{xx}(X(t)),
\end{equation}
and hence determined by the solution $X(t)$ satisfying the uncoupled equation~(\ref{motion of nonphysical vacuum point}). 
System~(\ref{u1 eta2 system}) admits an immediate reduction: the first equation is of Riccati type, hence the substitution 
\begin{equation}
\label{phsub}
u_1={\dot{\phi} \over \phi}
\end{equation}
for a new dependent variable $\phi(t)$ allows the reduction of system~(\ref{u1 eta2 system})  to a single second order ODE,
\begin{equation}
\label{pheq}
\ddot{\phi} +{2 C\over \phi^2}+ 2\, b_2 \, \phi=0 \,, 
\end{equation}
since the $\dot \eta_2$ equation can be integrated at once, 
\begin{equation}
\dot\eta_2+{3 \dot \phi \over \phi}\eta_2=0 \quad \Rightarrow \quad \eta_2={C \over \phi^3} \,, 
\label{ph3}
\end{equation}
for some constant $C$.

\begin{rem}
\label{rmrk}
A few comments are now in order:
\begin{enumerate}
\item
A glance at Table~(\ref{table:1}) and at system~(\ref{hierarchy eta simplified})-(\ref{hierarchy u simplified}) shows that beyond~(\ref{u1 eta2 system})  the equations of the hierarchy constitute a recursive system of \textit{linear}  differential equations for the index-shifted pair of unknowns $\{u_n,\eta_{n+1}\}$. In fact, the condition $\eta_1=0$ eliminates the term containing $u_{n+1}$ from the summation~(\ref{hierarchy u simplified}) making the system closed at any order $n$ with respect to all the dependent variables up to that order, and,  more importantly, past the first equation pair the system is { linear}, as its coefficients are determined only by the lower index variables $u_i$, $\eta_j$, with $i<n$ and $j<n+1$. 
For this reason, the possible occurrence of movable singularities, determined by the initial conditions,  is entirely governed by the leading order nonlinear pair of equations (\ref{u1 eta2 system}). (For the present case of nonphysical vacuum, this extends to non-flat bottoms an analogous result in~\cite{Camassa-wetting-mechanism}).  Thus, the significance of system~(\ref{u1 eta2 system}) together with~(\ref{motion of nonphysical vacuum point}) has a `global' reach that extends well beyond that of just the first equations in the hierarchy, as it encapsulates the nonlinearity of the parent PDE, at least within the class of power series initial data with an initial nonphysical vacuum point considered here.
\item
If one assumes that $u$, $\eta$ and the bottom topography 
at the initial time are in fact (one-sided) analytic functions {admitting the expansion (\ref{sol-exp})}, 
then the Cauchy–Kovalevskaya Theorem (see, e.g. \cite{Evans98}) assures that  the solution of the initial value problem of system~(\ref{SWE wavefront coordinates}) exist 
and is analytic for some finite time determined by the initial interval of convergence, i.e., by the initial values of the series' coefficients. Depending on these initial data, the 
maximum time of existence could then be completely determined by the reduced system~(\ref{u1 eta2 system}) and~~\ref{b2t}), since, as per the previous comment, the 
infinite system of equations past the~$(u_1,\eta_2)$-pair is linear and so its singularities coincide with those of the forcing functions, which in turn are entirely determined by 
this pair.

%
%

\item
Equation~(\ref{pheq}) is isomorphic to that of a point (unit) mass subject to a force field $-(2 C/\phi^2 +b_2\phi)$, which in general will be time-dependent through $b_2$. This can have interesting consequences. For instance,  the coefficient $b_2$ can be time-periodic by the choice 
\begin{equation} 
\label{duff}
b(x)=c_0+{c_2 \over 2} x^2+{c_4 \over 4}x^4 \,,
\end{equation}
which makes equation~(\ref{motion of nonphysical vacuum point}) for $X(t)$ that of a Duffing oscillator~\cite{Verhulst}, 
\begin{equation} 
\label{duffx}
\ddot{X}+c_2 X+c_4X^3=0\,.
\end{equation}
As well known (see e.g.~\cite{Verhulst}) this equation has periodic solutions depending on the constants $c_1$ and $c_2$ and on its initial conditions. In this case, for some classes of initial data, equation~(\ref{pheq}) can be viewed as that of a parametrically forced nonlinear oscillator, and resonances due to the periodic forcing $b_2(t)=c_2+3 c_4 \big(X(t)\big)^2$ from the solutions of~(\ref{duffx}) could arise, which in turn could generate nontrivial dynamics. This would further enrich the types of time evolution of PDE solutions supported by this class of bottom profiles with nonphysical vacuum initial data. 
\item
Depending on initial conditions, bottom topographies with {\it local} minima, such as in the above case~(\ref{duff}) when $c_2>0$ and $c_4<0$, can lead to an overflow of part of an initially contained fluid into the downslope regions, whose evolution could develop singularities in finite times of the ODE (and so ultimately of the PDE) solutions as an initially connected fluid layer could become disconnected. 
\end{enumerate}
\end{rem}

Further aspects of the solution behaviour of the  nonphysical vacuum case, for the special case of a parabolic bottom with an upward concavity, will be characterized in Section~\ref{section: parabolic solutions}.
Next, we will switch our focus to the case of a physical vacuum, which is significantly different as some of the properties  of its nonphysical counterpart cease to hold.  

%

%

\subsection{Physical vacuum points}
\label{sub: physical dry points}
The case of ``physical'' vacuum boundaries is characterized by $\eta_1(0)\ne 0$. When this condition holds, equation (\ref{wavefront acceleration}) is no longer sufficient to determine the motion of the wavefront $X(t)$ and  the whole hierarchy of equations (\ref{hierarchy eta simplified})--(\ref{hierarchy u simplified}), for the unknowns $X(t),\{u_n(t),\eta_n(t)\}$, $n=0,1,\dots$ is now completely coupled. Indeed, at any given integer $n$, (\ref{hierarchy eta simplified})--(\ref{hierarchy u simplified}) involve variables of order $n+1$. Thus the wavefront approach appears to be less effective to study physical vacuum points. This issue was not present in the setup of Gurtin of 
Section~\ref{section: wavefront analysis}, where the wave propagated over a constant state background. Indeed, whether it was $\zeta_1=0$ or $\zeta_1\ne 0$, the wavefront expansion was always sufficient to determine the motion 
of the wavefront~$X(t)$. The reason for this discrepancy is the lack of information about the function $u_0(\tau)$, now treated as a variable. This is in contrast with the previous case of nonphysical vacuum: the continuity requirement on the solution at the wavefront implied that $u_0(\tau)=0$. However, the same argument cannot be invoked again here. Indeed, there is no preferential way to assign the velocity field in the dry region. Moreover, even if appeal to continuity were to be made to guide a possible choice, the resulting velocity field would necessarily become discontinuous across the vacuum boundary.

This result, already proved in \cite{Camassa-dambreak} (§ 3.2) for a horizontal bottom, is extended here to more general topography: 
if the water surface is analytic and transverse to the bottom at the vacuum point $x=X(t)$, that is $\eta_1\ne 0$, and the velocity field $u$ is initially continuous, then $u$ becomes discontinuous at $x=X(t)$ for any $t>0$.
To see this, consider the Taylor expansion of the velocity field for the wavefront  in the vacuum region $\xi>0$,
\begin{equation}
    \restr{u}{\xi>0}=u_0'(\tau)+u_1'(\tau)\xi+u_2'(\tau)\xi^2+\dots,\qquad u_k'(\tau)=\lim_{\xi\to 0^+}\frac{1}{k!}\frac{\partial^k u}{\partial \xi^k}(\xi,\tau),
\end{equation}
whose coefficients evolve in time according to
\begin{equation}
\label{hierarchy u dry}
    \dot u_n'+(n+1)b_{n+1}+\sum_{k=1}^n k \, u_k' u_{n+1-k}'=0. \tag{$\textnormal{II}'$}
\end{equation}
For $n=0$, (II) and (\ref{hierarchy u dry}) yield
\begin{equation}
\label{dry-wet 0th order equations}
    b_1+\eta_1+\dot u_0=0,\qquad b_1+\dot u_0'=0, 
\end{equation}
and these two equations together imply 
\begin{equation}
    \frac{d\llbracket u\rrbracket}{dt}=\dot u_0'-\dot u_0=\eta_1\ne 0,
\end{equation}
where $\llbracket u\rrbracket\equiv u_0'-u_0$ denotes the jump of the velocity field.
Thus,  the graph of the velocity jump $\llbracket u\rrbracket(t)$ is never tangent to the curve $\llbracket u\rrbracket=0$. Moreover, since $\eta_1$ has constant sign (it is negative in the situation considered so far), the velocity jump  $\llbracket u\rrbracket(t)$ can vanish only once in a simple zero. Hence, adjusting for a possible time shift, we see that the discontinuity of the velocity field at the vacuum/dry point has to emerge at $t=0^+$.
Another way of stating this result is that a shock wave always forms at a physical vacuum point for the velocity-like component of the system. However, as pointed out in~\cite{Camassa-dambreak}, this is a non-standard kind of shock. Besides being uncoupled to the other dependent variable (in this case the water surface that maintains its initial continuity for a nonzero time interval) it also does not involve dissipation of conserved quantities, in general. It is indeed not too difficult to verify that the Rankine-Hugoniot conditions for mass, momentum and energy are all satisfied at the same time for this non-standard shock.

\section{Parabolic solutions}
\label{section: parabolic solutions}


As mentioned in Remark~\ref{rmgen} above, if the graph of the bottom topography is a polynomial of at most second degree, the shallow water equations 
(\ref{SWE}) 
admit a special class of explicit solutions which can be obtained by an exact truncation of the 
hierarchy~~(\ref{hierarchy eta simplified})-(\ref{hierarchy u simplified}). This result can be cast in different light by working directly with~(\ref{SWE}) and seeking solutions by self-similarity. With a quadratic bottom topography, similarity quickly leads to the ansatz of a quadratic form for~$\eta$ and a linear one for~$u$, as direct inspection shows that these forms would be maintained by the evolution governed by~(\ref{SWE}), provided the polynomial coefficients evolve appropriately in time.   With this in mind, it is convenient to seek solutions of~(\ref{SWE}) in the form 
\begin{figure}
    \centering 
    \includegraphics[width=0.6\textwidth]{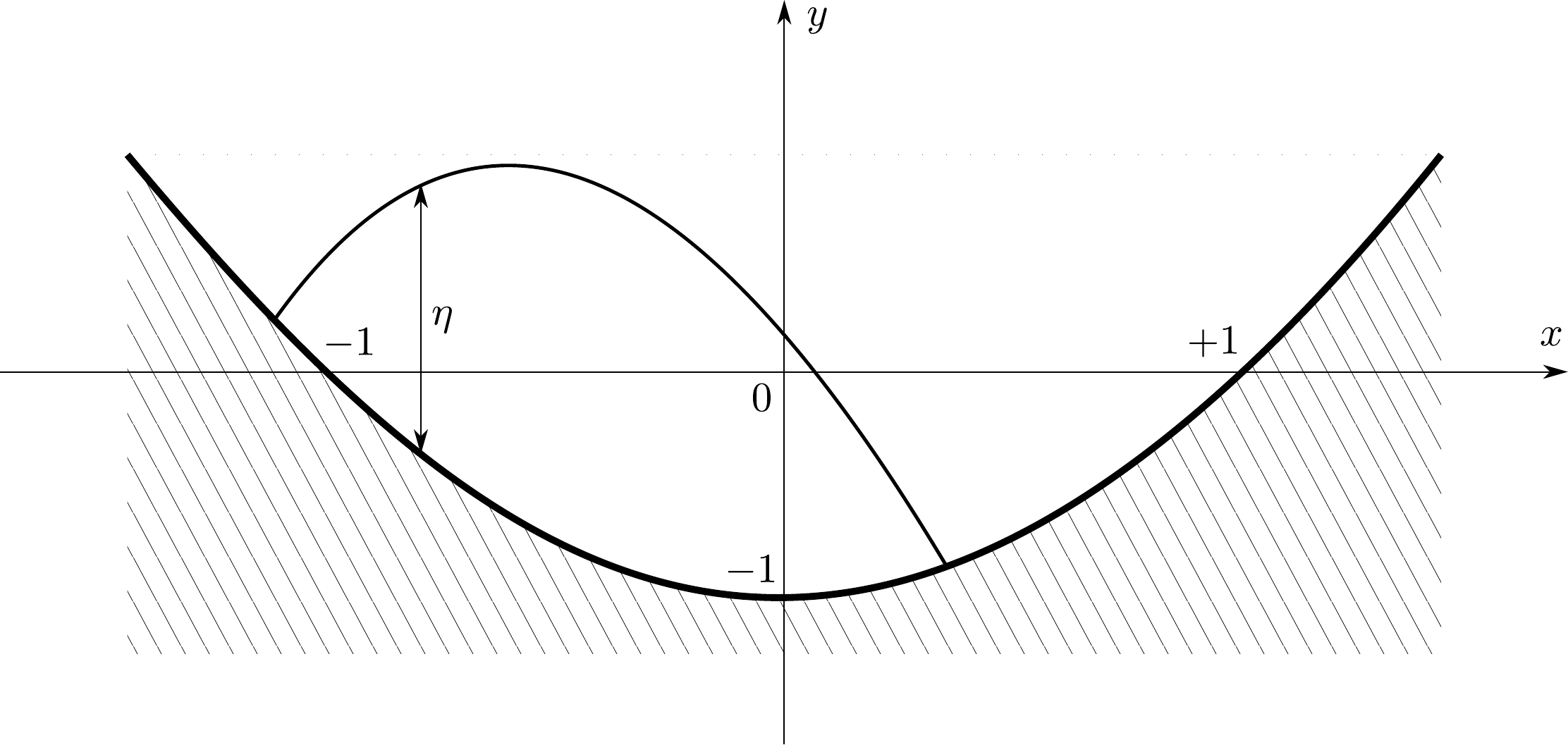}
    \caption{Schematic of the parabolic solution (\ref{5-field solution}) for $\gamma_0<0$. The water surface intersects the bottom at a couple of distinct dry/{physical}-vacuum points and the fluid domain is compact.}
    \label{fig:downward parabola}
\end{figure}
\begin{equation}
    \eta(x,t)=\mu(t)+\gamma(t)\big(x-\beta(t)\big)^2,\qquad u(x,t)=\delta(t)+\alpha(t)\big(x-\beta(t)\big).
    \label{5-field solution}
\end{equation}
Solution ansatz of this form were considered for the flat bottom case by Ovsyannikov \cite{Ovsy1979} and extended  by Thacker~\cite{Tha81} to a parabolic bottom and three-dimensions. 
Expression~(\ref{5-field solution}) is advantageous for the study of system's~(\ref{SWE}) behaviour in the vicinity of a dry point~\cite{Camassa-Geom,Camassa-wetting-mechanism,Camassa-dambreak,Camassa19}.
The  polynomial coefficients are chosen so that the location $x=\beta(t)$ is the critical point of the function $\eta(\:\cdot\:,t)$, whereas $\mu(t)$ is the corresponding critical magnitude, i.e., $\beta(t)$ and $\mu(t)$ are defined by
\begin{equation}
    \frac{\partial\eta}{\partial x}\bigg|_{x=\beta(t)}=0, \qquad \mu(t)=\eta(\beta(t),t).
  \end{equation}
Furthermore, a simple calculation~\cite{Camassa-Geom} shows that $x=\beta(t)$ coincides with the (horizontal)  position of the center of mass of the fluid layer. 
\begin{figure}[t]
    \centering
    \includegraphics[width=0.43\textwidth]{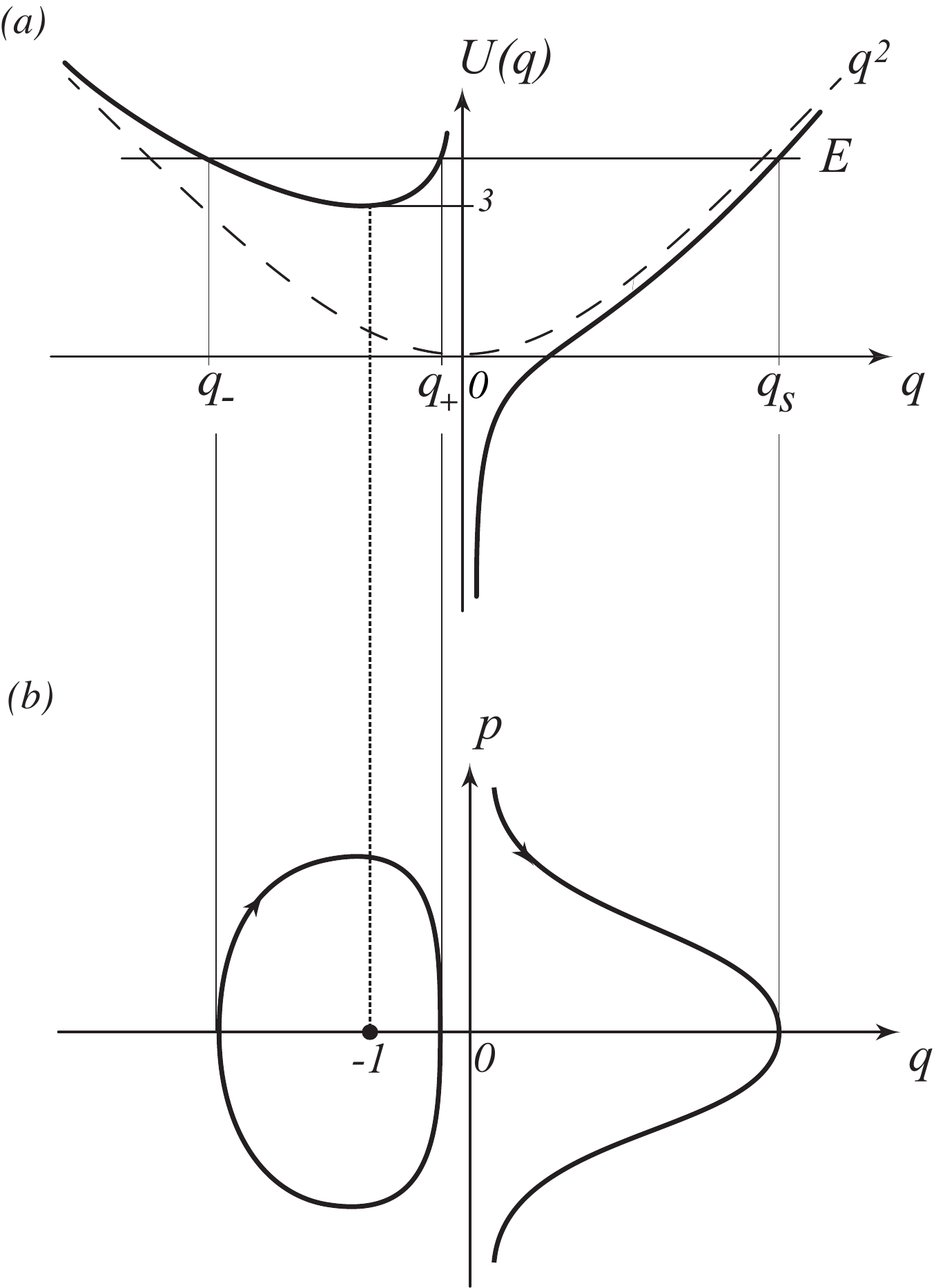}
    \caption{Sketch of the energy selection, $(a)$, and phase plane portrait, $(b)$, of the $(q,p)$ dynamics~(\ref{canon_sys}). There is only one (stable) equilibrium at $(q,p)=(-1,0)$. When $E>3$, orbits with $q<0$ ($\gamma<0$) are closed periodic cycles with $q\in[q_-(E),q_+(E)]$ around the equilibrium point, whereas orbits with $q>0$ 
($\gamma>0$) are open and attracted to the origin $q\to 0^+$ ($\gamma\to +\infty$), $q\in (0,q_s(E)]$. When $E<3$ only open orbits exist with $q\in (0,q_s(E)]$. Here $q_\pm(E)$ and $q_s(E)$ are, respectively, the two negative and the positive real roots $U(q)=E$ which exist for $E>3$. For $E<3$ only one real (positive) root exists. }
    \label{phase_portrait}
\end{figure}

The self-similar nature of these solutions, termed of the {\it second kind} in~\cite{Sedov}, is shown to be related to the scale invariance of system~(\ref{SWE}) in \cite{Camassa-Geom} (a complete account of symmetries and conservation laws for the shallow water system with a variable bottom can be found in \cite{Aksenov}). The relation between the form~(\ref{5-field solution}) and that of the truncated hierarchy to 
the~$(u_1,\eta_2)$ pair~(\ref{sol-exp}) for $u$ and $\eta$ is as follows (recall that $\gamma$ can be negative): 
\begin{gather}
\label{cnc}
X(t)=\beta(t)+\sqrt{{\mu(t)\over |\gamma(t)|}}\,, \qquad \eta_1(t)= -2\sqrt{|\gamma(t)| \mu(t)}\,,\qquad 
\eta_2(t)=\gamma(t)\,,
\\
u_0(t)=\delta(t)+\alpha(t)\sqrt{{\mu(t)\over |\gamma(t)|}}\,,\qquad u_1(t)=\alpha(t) \,.
\end{gather}
(Here, of the two physical-vacuum points $X(t)=\beta(t)\pm\sqrt{\mu(t)/|\gamma(t)|}$ we have chosen to work with the one to the right to conform with the definitions of Section~\ref{section:vacuum points}.)

%


For simplicity, in what follows the form 
\begin{equation}
    b(x)=x^2-1\,,
    \label{x2m1}
\end{equation}
for the bottom topography will be assumed, i.e., choose $c_0=-1$, $c_1=0$ and $c_2=1$ in~(\ref{quadb}). Plugging expressions~(\ref{5-field solution}) into the SWE~(\ref{SWE}) yields the following system of ODEs for
the time-dependent coefficients 
\begin{equation*}
    \dot\alpha+\alpha^2+2\gamma+2=0,\qquad \dot\gamma+3\alpha\gamma=0,\qquad \dot\mu+\alpha\mu=0,
\end{equation*}
\begin{equation}
\label{5-fields ODE new variables}
    \dot\beta=\delta,\qquad \dot \delta+2\beta=0 \,.
\end{equation}
These equations also follow directly from their series coefficient counterparts~(\ref{quadX}), of which they mirror the general structure, and from definitions~(\ref{cnc}). However, in this new set of variables the last two equations of~system (\ref{5-fields ODE new variables}) 
for the pair $\beta,\delta$ are uncoupled from the first three equations of the ODE for variables $\alpha$, 
$\gamma$ and $\mu$, where the nonlinearity of the system is concentrated.  The former are simply the equations of a harmonic oscillator in the variables 
$\beta$ and $\delta$, with $\delta$ then being the velocity~$\dot\beta$ of the center of mass. Thus, for this parabolic topography the center of mass $\beta$ oscillates harmonically about the origin with period $T=\pi\sqrt{2}$. Furthermore, the nonlinear behaviour of the system is entirely captured by the $\dot\alpha$ and $\dot\gamma$ equations, since the evolution of 
$\mu$ is slaved to that of $\alpha$. The possible occurrence of a finite-time singularity is thus governed by the coupled $(\alpha,\gamma)$ pair. It is easy to check that the quantity
\begin{equation}
\label{constant of the motion}
     H=\frac{\alpha^2-4\gamma+2}{2\gamma^{2/3}}
\end{equation}
is a constant of motion for system (\ref{5-fields ODE new variables}), and can be used to characterize the dynamics in the $(\alpha,\gamma)$-plane. 
%


As noted in \cite{Camassa-Geom}, the $H$ in~(\ref{constant of the motion}) is indeed a Hamiltonian function for the 
$(\alpha,\gamma)$ dynamics with respect to a noncanonical Poisson structure. With the change of variables 
\begin{equation}
q={1 \over \gamma^{1/3}}\,, \quad p={\alpha \over \gamma^{1/3}}
\label{qpdef}
\end{equation}  
the  $(\alpha,\gamma)$ dynamics is governed by the canonical system
\begin{equation}
 H={p^2 \over 2}+q^2-{2\over q}\equiv{p^2\over 2}+U(q)\, \quad \Rightarrow \quad
\dot{q}={\partial H \over \partial p}=p,\quad \dot{p}=-{\partial H \over \partial q} =-{d U \over d q}=-\left(2q+{2\over q^2}\right) \,,
\label{canon_sys}
\end{equation}
i.e., that of the one-dimensional dynamics of a point mass subject to a  (conservative) force with potential~$U(q)$. Note that this yields the same equation as~(\ref{pheq}) when $b_2=2$ and $C=1$. The analysis of such systems is straightforward, and we summarize here the main points.  The phase portrait of system~(\ref{canon_sys}) is depicted in Figure~\ref{phase_portrait}. 
The vertical asymptote at $q=0$ separates the  $(q,p)$ phase plane into two regions $q<0$ and $q>0$, corresponding 
to the curvature $\gamma$'s sign,  which is hence preserved by the time evolution. The initial data $\alpha(0)=\alpha_0$ and
$\gamma(0)=\gamma_0$ for the first pair of ODEs
in~(\ref{5-fields ODE new variables}) select the constant, $E$ say, of the corresponding energy level set, i.e.,  $H(q,p)=E$, and if $E>\min_{q<0}U(q)=3$, the value of $U$ at the stationary point $U'(q)=0$ for $q=-1$, two different evolutions are possible: bounded periodic orbits for $q<0$ (and hence $\gamma_0<0$) and open orbits with asymptote $q\to 0^+$, $p\to \infty$, 
 for $q>0$ (and hence $\gamma_0>0$ and $\gamma\to +\infty$).  If $E<3$ only open orbits are possible for initial conditions with $q>0$ (i.e., $\gamma_0>0$). Close to the fixed point $q=\gamma=-1$ and $p=\alpha=0$ the periodic motion is approximately harmonic with frequency $\sqrt{U''(-1)}=\sqrt{6}$, while for generic admissible energy levels $E$ the period is 
\begin{equation}
T'=2\int_{q_-(E)}^{q_+(E)}{dq \over {\sqrt{2(E-U(q))}}}\,, 
\label{prd}
\end{equation}
where $q_-(E)<q_+(E)<0$ are the two (negative) roots of $U(q)=E$.

In general, the solution $q(t)$ is expressed implicitly by 
\begin{equation}
t-t_0= \pm \int_{q(E)}^q {dq' \over {\sqrt{2(E-U(q'))}}}\,, 
\label{implc}
\end{equation}
and constructed by quadratures in terms of elliptic functions in the various cases picked by initial data, since the potential $U(q)$ leads to a cubic equation for the roots of $U(q)=E$, with the appropriate interpretation of the branch choice and the ``base" root $q(E)$. Our ultimate aim is to reconstruct the time dependence of the curvature function $\gamma(t)$, and it is convenient to change variable in the integrand to fix the location of the base root making it independent of the initial data. Further, 
setting $t_0=0$ and $p(0)=0$, the energy level $E$ can be expressed as a function of  the 
initial curvature $\gamma(0)=\gamma_0$, 
\begin{equation}
    E=(1-2\gamma_0)/\gamma_0^{2/3},
\end{equation}
a relation that can be made one-to-one  by confining $\gamma_0$ to the interval $(-1,0)$. 
Thus, all periodic orbits can be parametrized by an initial condition $\gamma_0\in(-1,0)$
with $\alpha_0=0$.
Changing variables by 
\begin{equation}
q={1\over \gamma_0^{1/3}\sigma} \Rightarrow \sigma=\left({\gamma \over \gamma_0}\right)^{1\over 3}
\label{sigmdef}
\end{equation}
the implicit form of the solution becomes 
\begin{equation}
\label{tau evolution}
    t=\pm\int_1^\sigma{d s\over \sqrt{4\gamma_0 s^5+2(1-2\gamma_0)s^4-2s^2}}
    =\pm\int_1^\sigma{d s \over s\sqrt{4\gamma_0(s-1)(s-\sigma_+)(s-\sigma_-)}} \,, 
\end{equation}
with the roots $\sigma_\pm$ given by
\begin{equation}
\label{tau pm}
    \sigma_\pm=\frac{-1\pm \displaystyle
   \sqrt{1-8\gamma_0}}{4\gamma_0}.
\end{equation}
\begin{figure}[t]
    \centering
    \includegraphics[width=0.4\textwidth]{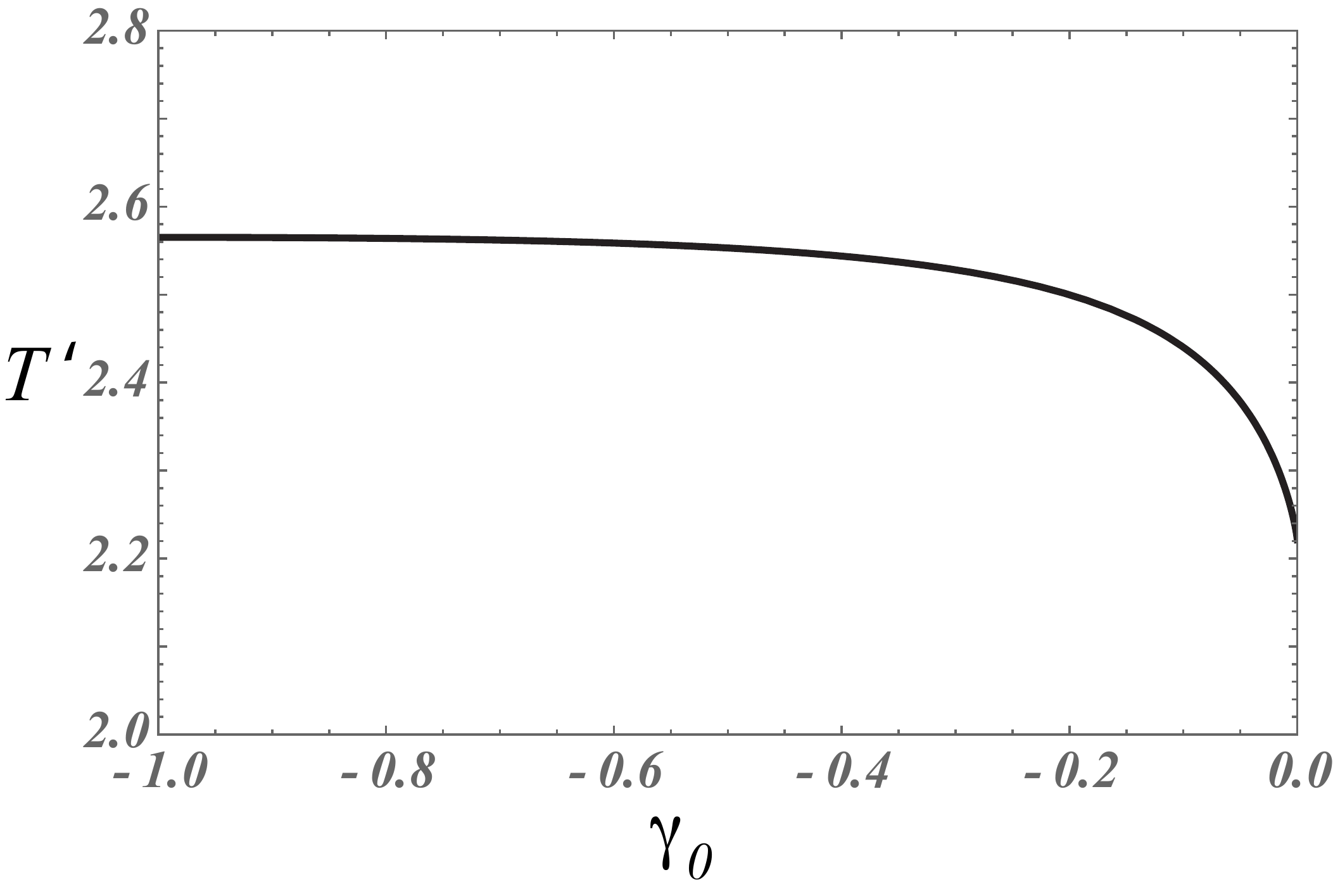}
    \caption{
    Graph of the period $T'$~(\ref{period elliptic}) as function of $\gamma_0$. The small amplitude oscillations, corresponding to an almost flat water surface with $\gamma_0\approx -1$, are quasi-isochronous with 
    period~$\pi\sqrt{2/3}\simeq 2.56$. For larger amplitudes, corresponding to orbits selected by the limit $\gamma_0\to 0$, the period decreases to the lower bound value $\pi/\sqrt{2}\simeq 2.22$.}
    \label{fig period}
\end{figure}

The quadrature expression for~(\ref{tau evolution}) in terms of elliptic functions is 
\begin{equation}
    t=\pm\frac{{g}}{2\sqrt{|\gamma_0|}}\left(\frac{1}{\sigma_+}F(\phi,k)+\Big(1-{1\over \sigma_+}\, \Big)\Pi(\phi,n,k)\right) \,.
 \label{ellptperiod}
\end{equation}
Here, the roots of the elliptic integral are such that  
$
\sigma_+<1<\sigma_-
$
(and $\sigma_+$ is negative) when $\gamma_0<0$, while for $\gamma_0>0$ the roots $\sigma_\pm$ are either real or complex conjugate with negative real part;   
$F(\phi,k)$ and $\Pi(n;\phi,k)$ are the incomplete elliptic integrals of the first and third kind~\cite{Byrd}, respectively, and their arguments are 
\begin{gather}
    g=\frac{2}{\sqrt{\sigma_--\sigma_+}},\qquad k=\sqrt{\frac{\sigma_--1}{\sigma_--\sigma_+}},\qquad n=\sigma_+ k^2\,, \qquad 
    \sin \phi=\sqrt{\frac{1}{k}\frac{\sigma-1}{\sigma-\sigma_+}}\,.
\end{gather}
%
%

\subsection{The sloshing solution}
\label{sub: gamma_0<0}
When $\gamma_0<0$, the physical system consists of a water ``drop" of finite mass, sloshing within a parabolic-shaped bottom (see Figure \ref{fig:downward parabola}). The fluid domain is compact at all times, and is bounded by two physical vacuum points. The solution is globally defined and no singularities develop in time. As already noticed, the motion in the $(\alpha,\gamma)$-plane is periodic with period $T'$, given by~(\ref{prd}), whose quadrature expression in terms of elliptic integrals is given by~(\ref{ellptperiod}) by setting  $\sigma=\sigma_-$,
\begin{equation}
\label{period elliptic}
    T'=\frac{2}{\sqrt{|\gamma_0|}\sqrt{\sigma_--\sigma_+}}\left(\frac{1}{\sigma_+}F\left({\pi \over 2},k\right)+\Big(1-{1\over \sigma_+}\, \Big)\Pi\left({\pi \over 2},n,k\right)\right).
\end{equation}
As a function of $\gamma_0$, the graph of the period $T'$ is depicted in Figure~\ref{fig period}, where we can see that $T'$ lies in the interval $T'\in (\pi/\sqrt{2},\pi\sqrt{2/3})$. The five dependent variables which parametrize the self-similar solutions are naturally arranged in the triplet $(\alpha,\gamma,\mu)$ and doublet $(\beta,\delta)$, whose dynamics can be viewed as representing that of a two (uncoupled) degree-of-freedom mechanical system. From this viewpoint, as mentioned in~Remark~\ref{rmrk}, the self-similar solutions can be interpreted as a reduction of the infinite number of degrees of freedom, or modes, of the wave-fluid continuum to just two. Only one of these degrees of freedom (that corresponding to the triplet~$(\alpha,\gamma,\mu)$) captures the nonlinearity of the original PDE, with the period of oscillation being a function of the mechanical system's initial conditions (and hence energy). Thus, for generic initial data the nonlinear period $T'$ given by~(\ref{period elliptic}) cannot be expected to be a rational multiple of that of the linear center-of-mass oscillator, $T=\pi \sqrt{2}$, and the oscillations of the fluid system will be quasiperiodic. Of course, in general the fluid continuum admits infinitely many configurations sharing the same position of the center of mass, and this lack of periodicity can be seen as a ``legacy," in this simple setting of self-similar solutions, of the original continuum dynamics. Nonetheless, given the initial condition dependence of the period $T'$ and its consequence on the continuum of values that this can attain, we can expect to  have infinitely many initial configurations with periodic evolutions when $T'=m T/n$ for some integers $m$ and $n$. 

\begin{figure}
    \centering
    \includegraphics[width=0.8\textwidth]{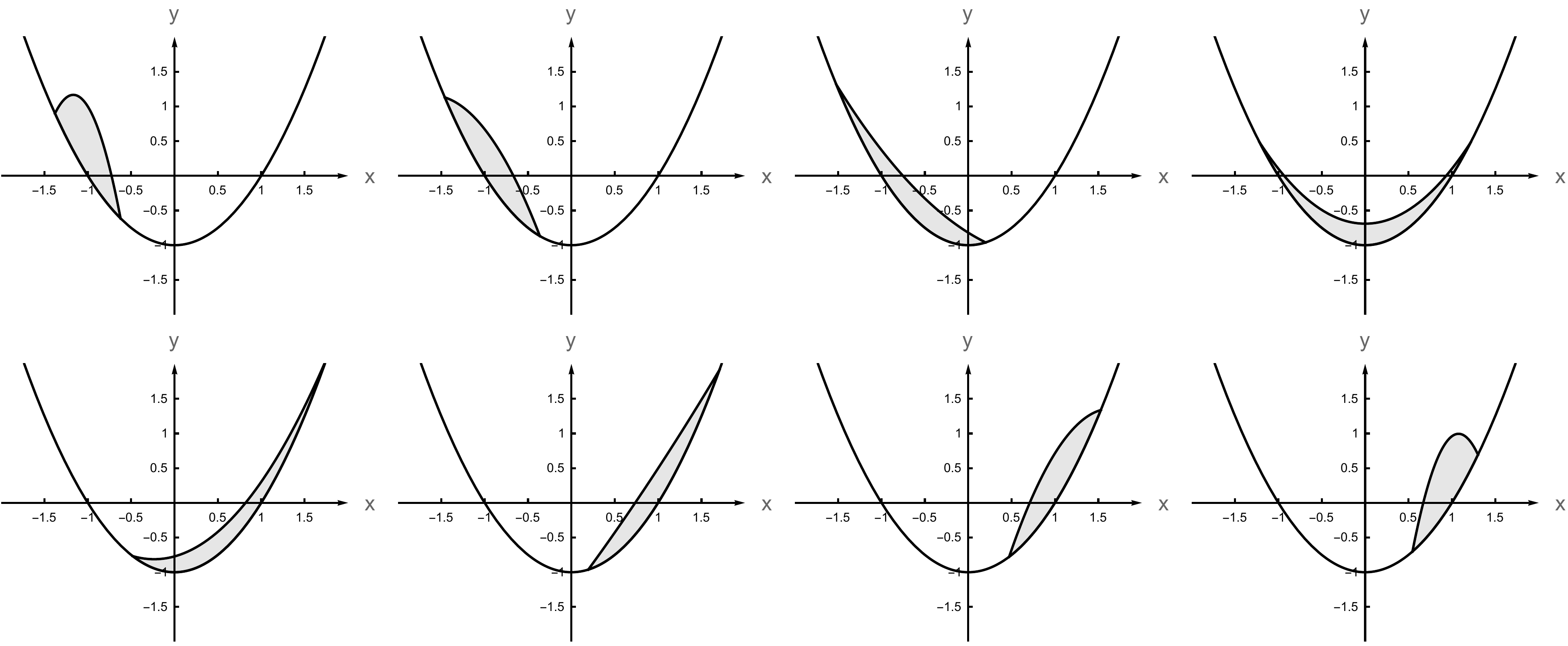}
    \caption{Sample motion of a water mass sloshing inside a parabolic-shaped bottom. The snapshots depict one oscillation of the curvature, that is $0\le t\le T'$. The motion is started from rest, that is $\alpha_0=0,\delta_0=0$, with a water surface specified by $\gamma_0=-7,\mu_0=1,\beta_0=-1$. The snapshots correspond to times $t=0, 0.30, 0.60, 1.11, 1.60, 2.00, 2.22, 2.50$. The second to last  frame corresponds 
    to half of the sloshing (center-of-mass) period $T$, which should be compared to the last frame taken at the time of (curvature) period $t=T'$ (specularly symmetric with respect to $y$-axis  to the frame at $t=0$).}
    \label{figure oscillating drop}
\end{figure}

The 
oscillatory motion of the full fluid layer is exemplified in Figure~\ref{figure oscillating drop} and in animations provided in the Supplementary Material (see also, among recent papers on the 
subject, reference~\cite{SES06} where the authors study the particular case of sloshing solutions where the free surface over a parabolic bottom is a straight-line segment at all times). At the fixed point $(\gamma_0,\alpha_0)=(-1,0)$, the free surface is a straight segment with slope 
$2\beta(t)$ oscillating sinusoidally, with the end points sliding along the parabolic boundary. 
In a neighborhood of this fixed point, the curvature $\gamma$ undergoes small amplitude oscillations with frequency $\displaystyle\sqrt{U''(-1)}=\sqrt{6}$, hence the upper bound for the period ~$T'$. The lower bound can be obtained explicitly by the asymptotic of large amplitude oscillations corresponding to $\gamma_0\to 0^-$; in fact, 
in this limit the integral in~(\ref{tau evolution}) becomes asymptotically
\begin{equation}
    T'\sim \sqrt{2} \int_1^\infty {\de s \over s \sqrt{s^2-1}}={\pi \over \sqrt{2}}
    \qquad \textnormal{as}\qquad \gamma_0\to 0^-.
\end{equation}
Note that the two periods $T$ and $T'$ of the center of mass and the water surface curvature achieve the lowest order rational relation in this  limit of large oscillations,  since we have
\begin{equation}
    T'\to\frac{T}{2}\qquad\textnormal{as}\qquad  \gamma_0\to 0^-.
    \label{subha}
\end{equation}
In fact, in this limit the free surface approaches a Dirac-delta function shape at the end points of 
the $\gamma$~oscillation, which necessarily makes the the center of mass coincide with the support of the delta function; this explains the subharmonic resonance expressed by~(\ref{subha}) as $\gamma_0\to 0^-$. Of course, some of these considerations have to be interpreted with an eye on the physical validity of the long wave model in the first place. When the evolution lies outside of the long-wave asymptotics used for the model's derivation,  some of the solutions presented here  
could  at most be expected to provide an illustration of trends in the actual dynamics of water layers. Note, however, that the robustness of the predicting power of these shallow water models can go beyond their strict asymptotic validity, as established by comparison with experiments and direct numerical simulations of the parent Euler system (see,.e.g.~\cite{Stoker48,Camassa19}).

\begin{figure}
    \centering
    \includegraphics[width=0.6\textwidth]{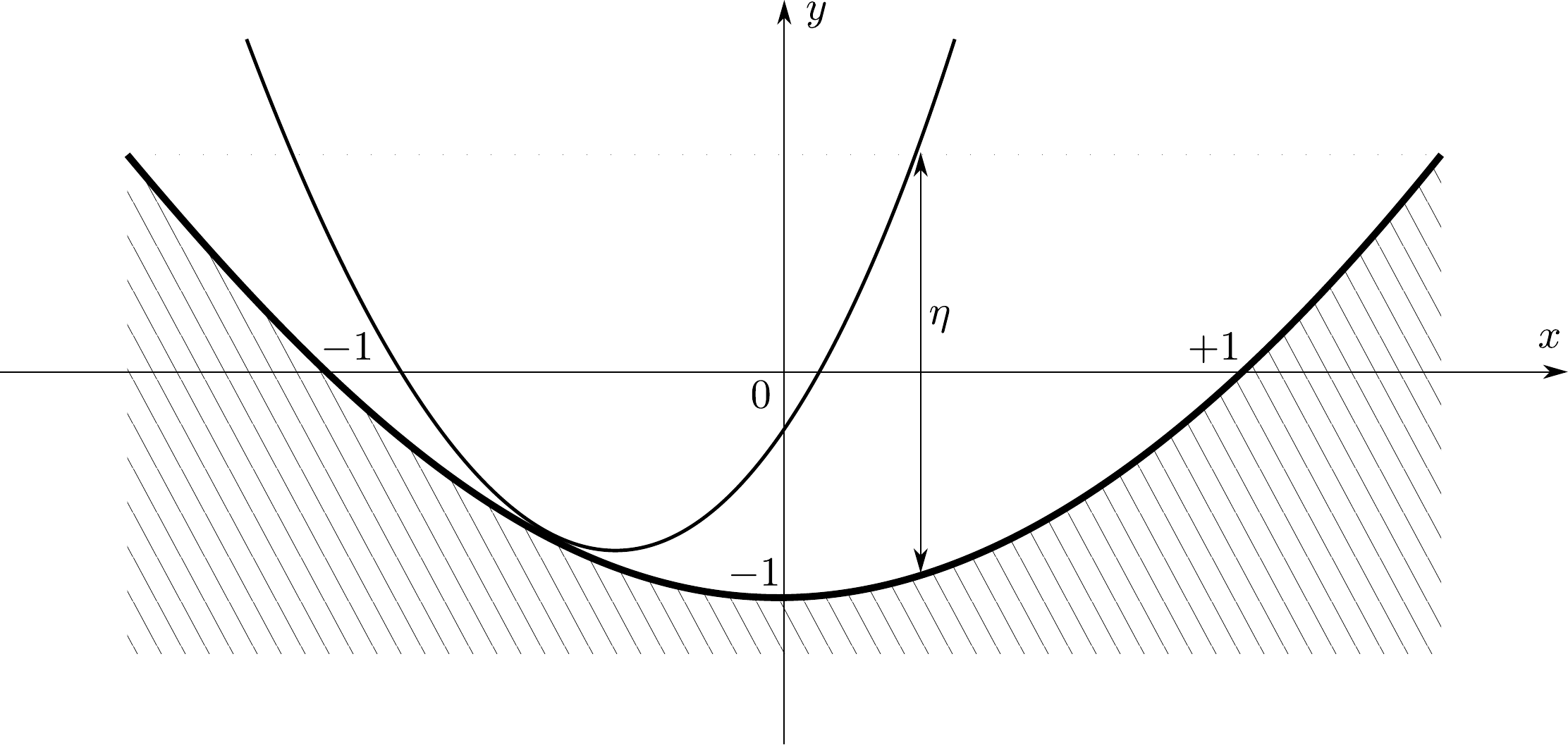}
    \caption{A schematic of the parabolic solution (\ref{5-field solution}) for $\gamma_0>0$ and $\mu_0=0$. The water surface is tangent to the bottom at a single nonphysical vacuum point, $x=\beta(t)$.}
    \label{fig:upward parabola}
\end{figure}

\subsection{The blow-up solution}
When $\gamma_0>0$, the parabolic water surface has positive curvature with magnitude larger than that of the bottom.  Given the definition~(\ref{5-field solution}) of the layer thickness $\eta(x,t)$, this means that whenever 
$\mu<0$ the possibility  of a dry region exists, i.e., the water surface intersects the bottom at a pair of points, which can merge in a limiting case
as shown in Figure~\ref{fig:upward parabola}. As in the previous section, the motion of the ``center of mass" $x=\beta(t)$ is still oscillatory, with period $T=\sqrt{2}\pi$, though the notion of mass for these unbounded solution needs to be generalized. On the other hand, the solution of system (\ref{5-fields ODE new variables}) is unbounded in the $\gamma,\alpha$ components. Viewed from the equivalent variables $q,p$, the interpretation of the dynamics set by the force potential $U(q)$ immediately shows that the blow-up of $\gamma(t)$ must occur in finite time. In fact, $U$ gives rise to an attractive ``gravity-like" force which diverges as $\sim -2q^{-2}$ in the limit 
$q\to 0^+$; this leads to $q\to 0^+$ and hence $\gamma(t)\to \infty$ in finite time starting from any finite initial condition $q_0>0$, $p_0=0$. Note that an ``escape velocity" does not exist in this gravitational force analogy, owing to the presence of the confining term $-q^2$ in the full expression of $U$, so that the ``plunge" into the origin $q=0$ will occur for any initial finite $p_0\neq 0$. Also note that the collision with the center of attraction at $q=0$ can be continued in time, by prolonging the trajectory so that $p$ jumps from $-\infty$ to $+\infty$, which effectively ``closes" the orbits in right half phase-plane $q>0$ and makes them degenerate oscillations, with an infinite jump in the momentum-like variable at $q=0$.  

\begin{figure}[t]
    \centering
    \includegraphics[width=0.4\textwidth]{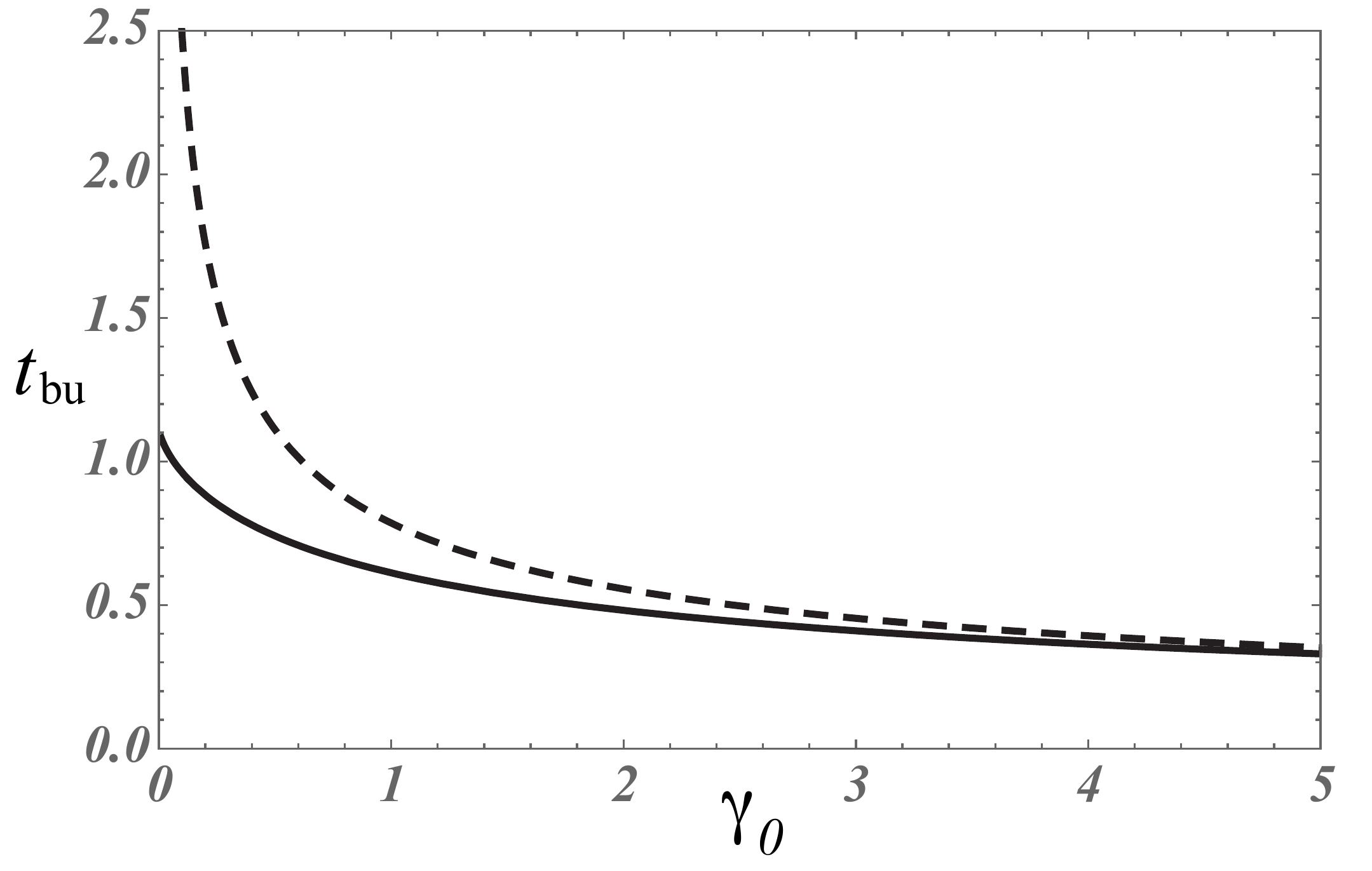}
    \caption{Graph of the blowup time~(\ref{blow-up time integral}) vs.~$\gamma_0$ for parabolic bottom $b(x)=x^2-1$ (solid), compared to its counterpart $\pi /(4\sqrt{\gamma_0})$ (dash) for the flat bottom case with positive initial~$\gamma_0$~(see \cite{Camassa19},~\S~2). 
    }    
    \label{tblows}
\end{figure}

Just as for the period of the oscillatory solutions, the blowup time counterpart  has a closed form expression in terms of elliptic functions, 
 \begin{equation}
\label{blow-up time integral}
    t_\textnormal{bu}\equiv\int_1^{+\infty}\frac{\de s}{s\sqrt{2(s-1)(2\gamma_0 s^2+s+1)}}=-\frac{1}{\sqrt{\gamma_0}\,\sigma_-
   \displaystyle{\sqrt{1-\sigma_-}}}\left\{-F\Big({\pi\over 2},k\Big)+\Pi\left(n;{\pi\over 2},k\right)\right\} \,, 
\end{equation}
which leads to the dependence on the initial curvature $\gamma_0$ shown in Figure~\ref{tblows}, where it is compared with the blowup time for the case of flat bottom discussed in~\cite{Camassa19}. Once again, limits $\gamma_0 \to 0^+$ and $\gamma_0 \to \infty$ can be analyzed, either directly from the integral or from known asymptotics of elliptic functions in~(\ref{blow-up time integral}), and it can be shown that 
$$
t_\textnormal{bu}\sim {\pi\over 2^{3/2}} \quad \textnormal{as}\quad \gamma_0\to0^+\,,
\qquad \textnormal{and} \qquad 
t_\textnormal{bu}\sim {\pi\over 4 \sqrt{\gamma_0}}\quad \textnormal{as}\quad\gamma_0\to \infty\,.
$$

\subsection{Piecewise parabolic solutions}
\label{section: piecewise parabolic solutions}
As an example of application of the methods described in Section~\ref{sub: piecewise initial conditions}, we consider here a particular class of piecewise initial conditions \cite{Camassa-wetting-mechanism,Camassa-dambreak,Camassa19},  whereby the fluid velocity is everywhere null, and the water surface is composed by a centered parabola continuously connected with a constant state on both sides (see figure~\ref{fig:pw parabola}):  
\begin{equation}
\label{pw parabolic initial conditions y}
    \zeta(x,0)=\begin{cases}0 &\textnormal{for }-1<x<-x_0\\ \zeta_{\mathrm{in}}(x) &\textnormal{for }-x_0\le x\le +x_0\\ 0 &\textnormal{for }+x_0<x<+1\end{cases},
\end{equation}
where
\begin{equation}
    \zeta_{\mathrm{in}}(x)=(\gamma_0+1)(x^2-x_0^2)\,.
\end{equation}
Due to the symmetry of this  configuration we can restrict our attention to the right semiaxis $x>0$. Immediately after the initial time, the corner at $x=x_0$ bifurcates into a couple of new derivative discontinuity points $X_\ell(t)$, $X_r(t)$, as described in Section~\ref{sub: piecewise initial conditions}. These points enclose the region which we called the \textit{shoulder} (see again Figure \ref{fig:spacetime partition} and \ref{fig:shoulder}). Equation (\ref{y_1(x)}), reported here for convenience,
\begin{equation}
\label{y_1(x) bis}
    \zeta_1(x)=\frac{(b(x_0)/b(x))^{3/4}}{\zeta_1(x_0)^{-1}+\tfrac{3}{2}(-b(x_0))^{3/4}I(x)},
\end{equation}
can be used to predict the onset of a shock at the wavefront $x=X_r(t)$. This equation describes the slope of the water surface $\zeta_1$ immediately behind the wavefront position, as it progresses towards the shoreline located at $x=1$. 
\begin{figure}[t]
    \centering
    \includegraphics[width=0.6\textwidth]{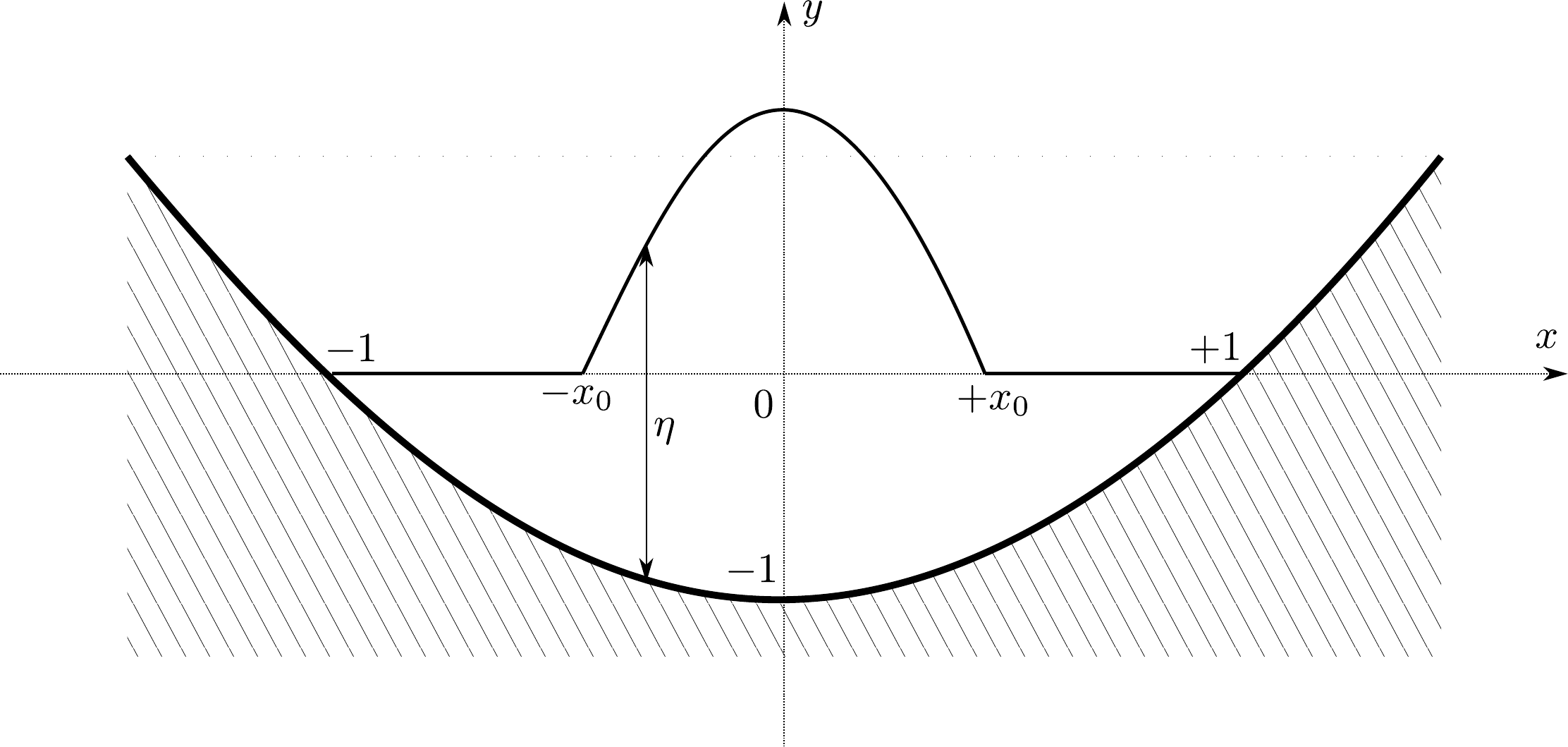}
    \caption{Sketch of piecewise initial conditions~(\ref{initial condition pw parabola}) for the water surface. A parabolic part is placed in the middle, and is continuously connected with a constant state at the sides through the singular points~$x=\pm x_0$. 
    }
    \label{fig:pw parabola}
\end{figure}
The appropriate initial condition (\ref{pw initial conditions y}), 
using~(\ref{shoulder initial slope}) in the shoulder region implies
\begin{equation} 
\label{initial condition pw parabola}
    \zeta_1(x_0)=\frac{\zeta_{\mathrm{in}}'(x_0)}{2} \,. 
\end{equation}
The integral $I(x)$ is defined in (\ref{I(x)}) and turns out to diverge when $x\to 1^-$, that is
\begin{equation}
\label{I(x) bis}
    I(x)=\int_{x_0}^x (1-\xi^2)^{-7/4}\de\xi\to+\infty\qquad \textnormal{as}\qquad x\to 1^-.
\end{equation}
Since $\zeta_1(x_0)$ in~(\ref{initial condition pw parabola}) is negative, the denominator of (\ref{y_1(x) bis}) will vanish for some $x_0<x_\textnormal{sh}<1$. This signifies that the wavefront will always break before reaching the shoreline. This result extends the one for a moving front of Gurtin~\cite{Gurtin75},  and holds for any regular bottom shape provided that the integral (\ref{I(x)}) diverges.

In this example the wavefront motion $X_r(t)$ can be obtained in closed form by solving (\ref{X_r X_ell}), which now reads
\begin{equation}
    \dot X_r=\sqrt{1-X_r^2},\qquad X_r(0)=x_0\, ,
\end{equation}
and has solution
\begin{equation}
    X_r(t)=\sin(\arcsin(x_0)+t)=x_0\cos t+\sqrt{1-x_0^2}\,\sin t.
\end{equation}
Note that $x=X_r(t)$ monotonically advances towards the shoreline $x=1$, and its speed $\dot X_r$ tends to zero for $X_r\to 1$.
The shock time for the wavefront can be computed as
\begin{equation}
\label{shock time}
    t_\textnormal{sh}=\arcsin(x_\textnormal{sh})-\arcsin(x_0).
\end{equation}
It is worth remarking that (\ref{shock time})
is an upper bound on the maximal interval of definition of the continuous solution arising from the initial conditions (\ref{pw parabolic initial conditions y}). Indeed, it is by no means certain for general initial data that the solution maintains smoothness up to 
$t_\textnormal{sh}$ and an earlier shock does not develop in the internal part of the shoulder region.  However, regardless of the details of the initial data, the above argument assures that a shock will always develop at the wavefront location  for $t=t_\textnormal{sh}$.

Next, we focus on the asymptotic behaviour of the shock location $x_\textnormal{sh}$ when $x_0\to 1^-$. We consider a sequence of initial conditions with the singular point 
$x_0$ approaching the shoreline $x=1$. Throughout this limit procedure, we allow for a sequence of initial values $\gamma_0$, $\zeta_{\mathrm{in}}(0)$ 
such that the initial slope $
\zeta_1(x_0)$ of the parabola, given by (\ref{initial condition pw parabola}), is held constant. If $x_0\to 1^-$ (and consequently $x\to 1^-$) we can derive an asymptotic 
estimate for the integral (\ref{I(x) bis}): with the shifted variable $s=1-\xi$, which tends to zero in this limit, integral (\ref{I(x) bis}) can be approximated by
\begin{equation}
\label{I(x) estimate}
    I(x)=\int_{1-x_0}^{1-x}\frac{\de s}{[s(2-s)]^{7/4}}\sim \frac{2^{1/4}}{3}\left\{(1-x)^{-3/4}-(1-x_0)^{-3/4}\right\}.
\end{equation}
The asymptotic estimate for the shock position $x_\textnormal{sh}$ is obtained by replacing $I(x)$ with this estimate. Thus, for $x_0\to 1^-$, we have
\begin{equation}
\label{interm-eq}
    (1-x_\textnormal{sh})^{-3/4}-(1-x_0)^{-3/4}+2^{3/4}\zeta_1(x_0)^{-1}(-b(x_0))^{-3/4}\sim 0.
\end{equation}
The bottom topography is approximately 
\begin{equation}
    -b(x_0)=(1-x_0^2)\sim 2(1-x_0) \qquad\textnormal{as}\qquad x_0\to 1^-.
\end{equation}
Equation~(\ref{interm-eq}) then simplifies to
\begin{equation}
    1-x_\textnormal{sh}-(1-\zeta_1(x_0)^{-1})^{-4/3}(1-x_0)\sim 0 \qquad\textnormal{as}\qquad x_0\to 1^-.
\end{equation}
Finally, as $\zeta_1(x_0)$ is constant throughout the limit procedure, it follows that 
\begin{equation}
\label{asymptotic shock position}
    (1-x_\textnormal{sh})\sim (1-x_0) \qquad\textnormal{as}\qquad x_0\to 1^-.
\end{equation}
Hence the shock position approaches the value $1$ at the same asymptotic rate as $x_0$.

According to (\ref{shock time}) and (\ref{asymptotic shock position}), the shock time $t_\textnormal{sh}$ tends to zero in the limit $x_0\to 1^-$. On the other hand, the limit configuration of (\ref{initial condition pw parabola}) corresponds to the situation already encountered in Section~\ref{sub: gamma_0<0}, where the parabolic surface intersects the bottom at $x=\pm x_0$ and the constant parts are not present. This observation suggests that this limiting  initial configuration gives rise to a shock immediately after the initial time $t=0$. This is a reasonable conclusion if interpreted in light of Section~\ref{sub: physical dry points}. Indeed, as seen in that section, the velocity field becomes discontinuous immediately after the initial time. On the other hand, as noted in Section~\ref{sub: gamma_0<0}, the water surface does not suffer from the same discontinuity, and for this reason the shock that arises has to be regarded as non-standard~\cite{Camassa-dambreak}. This phenomenon is related to the mixed character of the shallow water system, which is locally parabolic at a dry point. These nonstandard shocks do not share many of the classic properties of their fully hyperbolic counterparts. For example, a whole infinite family of  conserved quantities are also conserved across these shocks~\cite{Camassa-dambreak} (with the possible exception of the $u$-conservation only).


\FloatBarrier
\section{Conclusions and future directions}
\label{conc}

We have discussed several aspects of non-smooth wave front propagation in the presence of bottom topography, including the extreme case of vacuum/dry contact points, within the models afforded by long wave asymptotics and their hyperbolic mathematical structure. A notable extension of the flat bottom results in our previous work has been derived, in particular the finite time global catastrophe formation when the relative curvature of the interface with respect to that of the bottom is sufficiently large.  A detailed characterization of the evolution of initial data singularities has been presented; in particular, in the case of vacuum points and quadratic bottom topographies, we have illustrated this time evolution by closed form expressions derived by identifying a class of exact self-similar solutions.
As in the flat bottom case~\cite{Camassa-wetting-mechanism},
these self-similar exact solutions acquire a more general interpretation for entire classes of initial data, those that can be piecewise represented by analytic functions and admit nonphysical vacuum dry points, whereby the surface is tangent to the bottom at the contact point(s). In this case, a reduction of the PDE to a finite number of degrees of freedom mechanical system is possible, and this captures the whole nonlinear behaviour in the hierarchy of equations for the power-series coefficient evolutions. An interesting consequence of this analysis, which we will pursue in future work, is the classification of the dynamics, within the reduced dynamical system, for polynomial bottom profiles of order higher than quadratic. These profiles inject time dependent drivers in the reduced system,  which can lead to resonances and thus has implications for integrability of the full PDE evolution. Also currently under investigation is a notable application of our results to illustrate the transition from the nonphysical vacuum regime to the physical one, which does not appear to be fully understood~\cite{Serre2015} yet. The opposite  transition form physical to nonphysical vacuum regimes seems to be more approachable as exemplified in~\cite{Camassa-dambreak}, where the authors find the explicit local behavior of the transition in the flat bottom case.}

On the more physical level, much remains to be done to include other relevant effects. First, the presence of surface tension and its dispersive mathematical features can naturally be expected to play a significant role in regularizing the singular behaviours we have encountered. In particular, it is of interest to examine the precise role of dispersive dynamics vis-\`a-vis  the regions dominated by hyperbolic regimes such as the ``shoulders" we 
have identified in Section~\ref{sub: piecewise initial conditions}. In particular, the transition between hyperbolic and dispersive dominated dynamics, which can be expected to occur when the typical Froude number of the flow~$|u|/\sqrt{g |\eta|}$  nears unity (see, e.g.,~\cite{Wetal19}), would modify the shoulders' generation and evolution, which could require developing an intermediate model to study.    Further extension to stratified systems with two or more layers of different densities is also of interest, and in the special case of two-layer fluids in Boussinesq approximation our results can be applied almost directly thanks to the map to the shallow water system (paying attention to the additional subtleties caused by the map's multivaluedness)~\cite{EslerPierce2011,Ovsy1979}.

Finally, we remark that our investigation within 1-D models can be generalized to higher dimensions where, however, even for a flat bottom topography Riemann invariants may not 
exist, and hence some of the progress would have to rely on numerical assistance for extracting detailed predictions from the models. This would extend the two-dimensional topography results in~\cite{Tha81} by connecting self-similar ``core" solutions  to background states via the analog of ``shoulder" simple waves, similar to what we have done in our work here and elsewhere. Further, the complete analysis in three-dimensional settings should include classifications and time-estimates of gradient catastrophes. Thus, a priori estimates on the time of gradient blowup~\cite{Lax64} for hyperbolic one-dimensional systems 
should be extended to higher dimensions, with the core blowup of the self-similar solutions providing an upper bound for more general initial conditions. These and other issues will be pursued in future work.


\FloatBarrier
\section*{Acknowledgements}
We thank the anonymous referees for useful suggestions on the paper's exposition  and for alerting us to references~\cite{Tha81,SES06,Wetal19}.
This project has received funding from the European Union's Horizon 2020 research and innovation programme under the Marie Sk{\l}odowska-Curie grant no 778010
 {\em IPaDEGAN}. We also gratefully acknowledge the auspices of the GNFM Section of INdAM under which part of this work was carried out. RC thanks the support by 
 the National Science Foundation under grants RTG DMS-0943851, CMG ARC-1025523, DMS-1009750, DMS-1517879, DMS-1910824, and by the Office of Naval 
 Research under grants N00014-18-1-2490 and  DURIP N00014-12-1-0749. MP thanks  the   Department of Mathematics and its Applications of the  University of 
 Milano-Bicocca for its hospitality.

\setcounter{section}{0}
\renewcommand{\thesection}{\Alph{section}}

\section*{Appendix A: {Shallow water equation with variable bottom}}
\renewcommand{\theequation}{A.\arabic{equation}}
\renewcommand{\thefigure}{A.\arabic{figure}}
%
\label{app-SWE}
\setcounter{equation}{0}
\setcounter{figure}{0}
\setcounter{proposition}{0}

For the reader's convenience, we list here the different forms of the shallow water system analyzed in this paper (see \cite{Carrier58,Stoker48}). Following \cite{Carrier58}, we use a ``$*$'' superscript to denote a dimensional quantity; symbols not accompanied by this marker will be reserved for nondimensional ones. The one-dimensional shallow water equations (SWE) with general {smooth} bottom topography can be written as
\begin{equation}
\label{dimensional shallow water equations}
    \eta^*_{t^*}+(u^*\eta^*)_{x^*}=0,\qquad u^*_{t^*}+u^* u^*_{x^*}+g(\eta^*+b^*)_{x^*}=0,
\end{equation}
where $\eta^*(x^*,t^*)$ represents the thickness of the water layer, $b^*(x^*)$ is the bottom elevation over a reference level, and $u^*(x^*,t^*)$ represents the layer-averaged horizontal component of the velocity field. As usual, $g$ stands for the constant gravitational acceleration. An alternative form of  system (\ref{dimensional shallow water equations}) is obtained by replacing the layer thickness $\eta^*$ with the elevation of the water surface
\begin{equation}
\zeta^*(x^*,t^*)=b^*(x^*)+\eta^*(x^*,t^*),  
\end{equation}
which turns the SWE into
\begin{equation}
\label{dimensional shallow water equations 2nd form}
    \zeta^*_{t^*}+[u^*(\zeta^*-b^*)]_{x^*}=0,\qquad u^*_{t^*}+u^* u^*_{x^*}+g \zeta^*_{x^*}=0.
\end{equation}
In this paper we will exclusively resort to the dimensionless form of (\ref{dimensional shallow water equations}) and (\ref{dimensional shallow water equations 2nd form}), 
\begin{equation}
    \label{dimensionless shallow water equations}
    \eta_t+(u\eta)_x=0,\qquad u_t+u u_x+\eta_x+b_x=0,
\end{equation}
and
\begin{equation}
\label{dimensionless shallow water equations y}
    \zeta_t+[u(\zeta-b)]_x=0,\qquad u_t+u u_x+\zeta_x=0.
\end{equation}
While the bottom term appears consistently with the long wave asymptotic approximation under which the shallow water equations are derived, it introduces additional parameters that can be scaled out in the dimensionless form  (see \cite{Carrier58}). For example, when the bottom is a parabola with upward concavity,
\begin{equation}
    b^*(x^*)=\kappa x^{*2}-Q, \qquad \kappa>0,\qquad Q>0,
\end{equation}
then a suitable scaling of the variables shows that SWE are independent of the parabola parameters. One appropriate choice is
\begin{gather}
\label{dimensionless variables}
    x=\frac{x^*}{l_0},\qquad t=\sqrt{g\kappa}\:t^*,\qquad u=\frac{u^*}{l_0\sqrt{g \kappa}},\qquad \eta=\frac{\eta^*}{\kappa l_0^2},\\ \zeta=\frac{\zeta^*}{\kappa l_0^2},\qquad b=\frac{b^*}{\kappa l_0^2}.
\end{gather}
Here, the characteristic length $l_0=\sqrt{Q/\kappa}$ is chosen so as to simplify the bottom term, which in the new variables reads
\begin{equation}
\label{dimensionless parabolic bottom}
    b(x)=x^2-1.
\end{equation}
We leave unspecified the bottom term in systems (\ref{dimensionless shallow water equations}) and (\ref{dimensionless shallow water equations y}) throughout Section~\ref{section: wavefront analysis} and Section~\ref{section:vacuum points}, whereas we consider the particular case (\ref{dimensionless parabolic bottom}) in 
Section~\ref{section: parabolic solutions}, where parabolic solutions are introduced.
\section*{Appendix B: Overview of near-front local analysis}
\renewcommand{\theequation}{B.\arabic{equation}}
\renewcommand{\thefigure}{B.\arabic{figure}}
\label{app-exp} 
\setcounter{equation}{0}
\setcounter{figure}{0}
\setcounter{proposition}{0}

We consider a globally continuous piecewise smooth solution to the SWE, with a jump discontinuity of the $k$-th order derivatives across a curve $x=X(t)$ in the spacetime plane. We further assume that a constant state is established on one side of this curve. The physical picture is that of a wavefront propagating over a still medium, whose trajectory in the $(x,t)$-plane is the curve $x=X(t)$ (see Figure \ref{fig:wavefront still fluid}). This is the classical problem setting investigated by Greenspan \cite{Greenspan58} and Gurtin \cite{Gurtin75}. It is convenient to make use of the form (\ref{dimensionless shallow water equations y}) of the SWE, so that the constant state on the right of the wavefront will simply be represented by $\zeta=0$, $u=0$.
There are many possible choices of variable sets adapted to the motion of the wavefront. The simplest one is a time-dependent space translation, which fixes to zero the wavefront position:
\begin{equation}
\label{space translation}
    \xi=x-X(t),\qquad \tau=t.
\end{equation}
Under the substitution (\ref{space translation}), the space and time derivatives transform as
\begin{equation}
    \partial_t=\partial_\tau-\dot X\partial_\xi,\qquad \partial_x=\partial_\xi,
\end{equation}
and the shallow water system (\ref{dimensionless shallow water equations y}) takes the form
\begin{equation}
\label{SWE wavefront coordinates y}
    \zeta_\tau-\dot X \zeta_\xi+[u(\zeta-b)]_\xi=0,\qquad u_\tau+(u-\dot X)u_\xi+\zeta_\xi=0.
\end{equation}
With the aim of extracting information on the system behaviour in a neighbourhood of the wavefront, we make the following ansatz on the form of the solution behind the wavefront, that is, for $\xi<0$,
\begin{align}
    \begin{split}
    \label{solution expansion y}
    &\restr{\zeta}{\xi<0}=\zeta_0(\tau)+\zeta_1(\tau)\xi+\zeta_2(\tau)\xi^2+\dots,\\ &\restr{u}{\xi<0}=u_0(\tau)+u_1(\tau)\xi+u_2(\tau)\xi^2+\dots.
    \end{split}
\end{align}
This ansatz essentially assumes that the solution is one-sided analytic in a neighbourhood of the wave front; henceforth, we will refer to it as the \textit{wavefront expansion} of the solution. Its coefficients are to be determined by substituting expansion~(\ref{solution expansion y}) back into (\ref{SWE wavefront coordinates y}). Although $\zeta_0$ and $u_0$ are formally included in these expansions, both their values are fixed to zero to ensure that the series (\ref{solution expansion y}) continuously connect to the constant state at $\xi=0$. Here $u_k(\tau)$ and $\zeta_k(\tau)$ can be viewed as limits of the relevant derivatives of $u$ and $\zeta$ for $\xi\to 0^-$, that is
\begin{equation}
    u_k(\tau)=\lim_{\xi\to 0^-}\frac{1}{k!}\frac{\partial^k u}{\partial \xi^k}(\xi,\tau),\qquad \zeta_k(\tau)=\lim_{\xi\to 0^-}\frac{1}{k!}\frac{\partial^k \zeta}{\partial \xi^k}(\xi,\tau).
\end{equation}
Similarly, the bottom topography is expanded in powers of $\xi$ (abusing notation a little 
by writing $b(\xi,\tau)=b(x)$),
\begin{equation}
\label{b expansion}
     b(\xi,\tau)=b_0(\tau)+b_1(\tau)\xi+b_2(\tau)\xi^2+\dots,
\end{equation}
where the time dependent coefficients $b_k$ are evaluated as
\begin{equation}
    b_k(\tau)=\frac{1}{k!}\frac{\partial^k {b}}{\partial \xi^k}(0,\tau)=\frac{1}{k!}\frac{d^k b}{d x^k}(X(\tau)).
\end{equation}
Plugging the expansions (\ref{solution expansion y}) and (\ref{b expansion}) into equations (\ref{SWE wavefront coordinates y}) and collecting the various powers of $\xi$ yields two infinite hierarchies of ODEs,
\begin{gather}
\label{hierarchy y}
\begin{split}
    &\dot \zeta_n+(n+1)\left[(u_0-\dot X)\zeta_{n+1}-b_0 u_{n+1}\right]-(n+1)b_{n+1}u_0+\cr
            &\hspace{5cm}+(n+1)\sum_{k=1}^n(\zeta_k-b_k)u_{n+1-k}=0,
    \end{split}
\end{gather}
\begin{gather}
    \dot u_n+(n+1)\left[\zeta_{n+1}+(u_0-\dot X)u_{n+1}\right]+\sum_{k=1}^n k \, u_k u_{n+1-k}=0
    \label{hierarchy u}
\end{gather}
for $n\geq 1$, while  $n=0$ yields
\begin{equation}
\label{0-0 system}
    -\dot X \zeta_1-b_0 u_1=0,\qquad \zeta_1-\dot X u_1=0 \,, 
\end{equation}
by recalling that $u_0=\zeta_0=0$.
The  structure of these equations determines the wavefront speed $\dot X$. This can be seen as follows: if the solution has discontinuous first derivatives across the wavefront then at least one among $u_1$ and $\zeta_1$ is different from zero; this in turn implies that the coefficient matrix of the linear system (\ref{0-0 system}) is singular, that is
\begin{equation}
\label{wavefront motion}
    \dot X^2+b_0=0.
\end{equation}
This first order nonlinear equation can in principle be solved to get the wavefront motion, which will depend solely on the bottom topography (with the choice of sign of $\dot X$ set by the initial data). The same result is obtained if the solution has a $k$-th order jump, with $k<\infty$, across the wavefront. For example, if the first derivatives were continuous and the second ones were not, we would obtain a linear algebraic system for $u_2$, $\zeta_2$ of the same form as (\ref{0-0 system}), once again reducing to (\ref{wavefront motion}). This is a general result (see e.g., \cite{Whitham99}); singularities of solutions of hyperbolic systems  propagate along characteristic curves.

The structure of equations (\ref{hierarchy y})--(\ref{hierarchy u}) is such that, for any fixed $n$,  variables of order up to $n+1$ are involved. However, higher order variables enter the system in such a way that both can be canceled by taking a single linear combination of the two equations. This becomes more transparent by writing system (\ref{hierarchy y})--(\ref{hierarchy u}) in matrix form
\begin{equation}
     \mathbf{\dot U}_n+(n+1)A \mathbf{U}_{n+1}+\mathbf{F}_n=0,\qquad A=\begin{pmatrix}-\dot X & -b_0\\ 1 & -\dot X\end{pmatrix},
\end{equation}
where $\mathbf{U}_n=(\zeta_n,u_n)^\top$, and $\mathbf{F}_n$ comprises variables of order up to $n$. The matrix $A$ is singular due to relation (\ref{wavefront motion}); thus, if the former equation is multiplied by the vector $(1,\dot X)$, the resulting scalar equation is free of higher order variables,
\begin{equation}
\label{linear combination}
    \dot \zeta_n+\dot X \dot u_n+(1,\dot X) \mathbf{F}_n=0.
\end{equation}
One of the two equations of order $(n-1)$ can now be used to replace $\zeta_n$ with $u_n$, or vice versa. For example, from (\ref{hierarchy u}) we get
\begin{equation}
\label{y_n to u_n}
    \zeta_n=\dot X u_n-\frac{1}{n}\bigg(\dot u_{n-1}+\sum_{k=1}^{n-1} k u_k u_{n-k} \bigg).
\end{equation}
Once (\ref{y_n to u_n}) is substituted into (\ref{linear combination}), a first order differential equation is eventually obtained, where neither $\zeta_n$ nor any higher order variable appear, and the only unknown is $u_n$. By iterating  this procedure, one can in principle solve the hierarchy (\ref{hierarchy y})--(\ref{hierarchy u}) up to any desired order.

The case $n=1$, for which (\ref{linear combination}) and (\ref{y_n to u_n}) give
\begin{equation}
    \dot \zeta_1+\dot X \dot u_1+2(\zeta_1-b_1)u_1+\dot X u_1^2=0, \qquad \zeta_1=\dot X u_1\, ,
    \label{ordo1}
\end{equation}
is of crucial importance for the prediction of gradient catastrophes at the wavefront.
The bottom term $b_1$ in (\ref{ordo1}) can be expressed in terms of $X(t)$. In fact, information about the bottom topography is encoded in the function $X(t)$, as obtained by solving (\ref{wavefront motion}). Taking the time derivative of (\ref{wavefront motion}) gives
\begin{equation}
\label{b1}
    2\ddot X+b_1=0.
\end{equation}
By eliminating $\zeta_1$ in~(\ref{ordo1}) with the second of (\ref{ordo1}), and using (\ref{b1}), we get
\begin{equation}
    \frac{\de}{\de t}(\dot X u_1)+\dot X \dot u_1+2(\dot X u_1+2\ddot X)u_1+\dot X u_1^2=0.
\end{equation}
The opposite substitution is also possible, and an equation for $\zeta_1$ can be obtained similarly. Upon rearranging the various terms, the evolution equation for $u_1$ and $\zeta_1$ take the form
\begin{equation}
\label{Riccati u_1 y_1}
    \dot u_1+\frac{3}{2}u_1^2+\frac{5\ddot X}{2\dot X}u_1=0,\qquad \dot \zeta_1+\frac{3}{2\dot X}\zeta_1^2+\frac{3\ddot X}{2\dot X}\zeta_1=0.
\end{equation}
 The Riccati-like mapping 
 $$
 \zeta_1 = \frac{2}{3} \dot{X}(t) \frac{\dot{\phi}(t)}{\phi(t)}
 $$ 
 linearizes  the second equation in~(\ref{Riccati u_1 y_1}), 
$$
5 \dot{\phi}(t) \ddot{X}(t)+2\ddot{\phi}(t) \dot{X}(t)=0\, , 
$$
which yields the quadrature solution
\begin{equation}
\label{y_1(t)}
    \zeta_1(t)=\frac{(\dot X(0)/\dot X(t))^{3/2}}{\zeta_1(0)^{-1}+\tfrac{3}{2}\dot X(0)^{3/2} I(t)},\quad \textnormal{where}\quad I(t)=\int_0^t\dot X(s)^{-5/2}\de s.
\end{equation}
It is often more convenient to focus on the dependence of the surface slope $\zeta_1$ on the wavefront position rather than on time. By making use of (\ref{wavefront motion}) and the change of variable $t=X^{-1}(x)$, the integral above can be expressed as
\begin{align}
\label{I(x)}
    I(x)=\int_{X^{-1}(x_0)}^{X^{-1}(x)}\dot X^{-\frac{5}{2}}(t')\de t'
    =\int_{x_0}^{x} \dot X^{-\frac{7}{2}}(X^{-1}(x'))\de x'
    =\int_{x_0}^{x} (-b(x'))^{-\frac{7}{4}}\de x',
\end{align}
where $x_0$ is the initial condition $X(0)=x_0$. Thus, equation (\ref{y_1(t)}) takes the form
\begin{equation}
\label{y_1(x)}
    \zeta_1(t)=\frac{(b(x_0)/b(x))^{3/4}}{\zeta_1(0)^{-1}+\tfrac{3}{2}(-b(x_0))^{3/4}I(X)} \Big{\vert}_{x=X(t)}\, .
\end{equation}
Of course, this construction relies on the existence of the inverse function $X^{-1}(x)$, and is well defined only for as long as the wavefront advances to the right.

The last equation allows one to predict the development of a gradient catastrophe at the wavefront location based on the given initial conditions and the bottom shape. It has been used by Greenspan \cite{Greenspan58} to investigate the breaking of a wavefront approaching a sloping beach with a straight bottom, and by Gurtin \cite{Gurtin75}, who extended Greenspan's analysis to general bottom shapes. In addition to the initial value $\zeta_1(0)$, the breaking conditions depend on the bottom topography, and in particular on its slope immediately close to the shoreline, which turns out to play a crucial role~\cite{Gurtin75} (see also \cite{Jeffrey80}).


\section*{Appendix C: {Flat bottom and piecewise parabolic solutions}}
\renewcommand{\theequation}{C.\arabic{equation}}
\renewcommand{\thefigure}{C.\arabic{figure}}
\setcounter{equation}{0}
\setcounter{figure}{0}
\setcounter{proposition}{0}
\label{flatbottomapp}
As remarked in Section~\ref{sec: reminder flat bottom}, the absence of the bottom term $b_x$ in the governing equations (\ref{dimensionless shallow water equations}) allows a thorough analysis of the solution to be performed thanks to the existence of Riemann invariants. Here, we apply the results  
of Section~\ref{sec: reminder flat bottom} to study the time evolution of initial conditions (\ref{pw parabolic initial conditions y}) in the special setting of a flat and horizontal bottom. The fluid is again assumed to be initially at rest, and the water surface is given by the piecewise parabolic initial conditions (\ref{pw parabolic initial conditions y}), which now read
\begin{equation}
\label{pw initial eta}
    \eta(x,0)=\begin{cases}
    Q & \textnormal{for } x<-x_0,\\
    \eta_{\mathrm{in}}(x) & \textnormal{for } -x_0\le x \le x_0,\\
    Q & \textnormal{for } x>x_0.
    \end{cases}
\end{equation}
Just like the previous section, the middle part is a centered parabola sector with downward concavity,
\begin{equation*}
    \eta_{\mathrm{in}}(x)=\gamma_0 (x^2-x_0^2)+Q,\qquad Q>0,\qquad \gamma_0<0.
\end{equation*}
The free surface continuously connects with the background constant state $\eta=Q$ at the  singular points
\begin{equation}
    x=\pm x_0
\end{equation}
where the first derivative $\eta_x$ suffers a jump discontinuity. Once again, give the symmetry of this initial configuration, we henceforth restrict our attention to $x>0$, as all the reasoning applies unchanged to $x<0$ as well. The singular point at $x=x_0$ bifurcates into a couple of new singular points after the initial time.  These are then transported along the characteristics $x=X_\ell(t)$ and $x=X_r(t)$. The positive one, $x=X_r(t)$, has uniform motion due to the presence of the constant state ahead of it. Specifically, we have
\begin{equation}
\label{X_r(t) flat bottom}
    x=X_r(t)=x_0+\sqrt{Q}\:t.
\end{equation}
On the other hand, the left wavefront, $x=X_\ell(t)$, exhibits a more complex behaviour, which depends on the parabolic part of the solution in the neighbouring region $x<X_\ell(t)$. This is given by
\begin{equation}
    \eta(x,t)=\gamma(t)x^2+\mu(t),\qquad u(x,t)=\alpha(t)x,
\end{equation}
and evolves in time according to the system of ordinary differential equations
\begin{equation}
    \dot\alpha+\alpha^2+2\gamma=0,\qquad \dot\gamma+\alpha\gamma=0,\qquad \dot\mu+\alpha\mu=0,
\end{equation}
with initial conditions 
$$
\alpha(0)=0, \qquad \mu(0)=\mu_0=Q-\gamma_0\, x_0^2\,.
$$
Therefore, $x=X_\ell(t)$ satisfies
\begin{equation}
\label{dX_ell/dt}
    \dot X_\ell=\alpha(t) X_\ell-\sqrt{\gamma(t) X_\ell^2+\mu(t)}\,,\qquad X_\ell(0)=x_0.
\end{equation}
\begin{figure}
    \centering
    ({\it a})
    \\
    \includegraphics[width=0.6\textwidth]{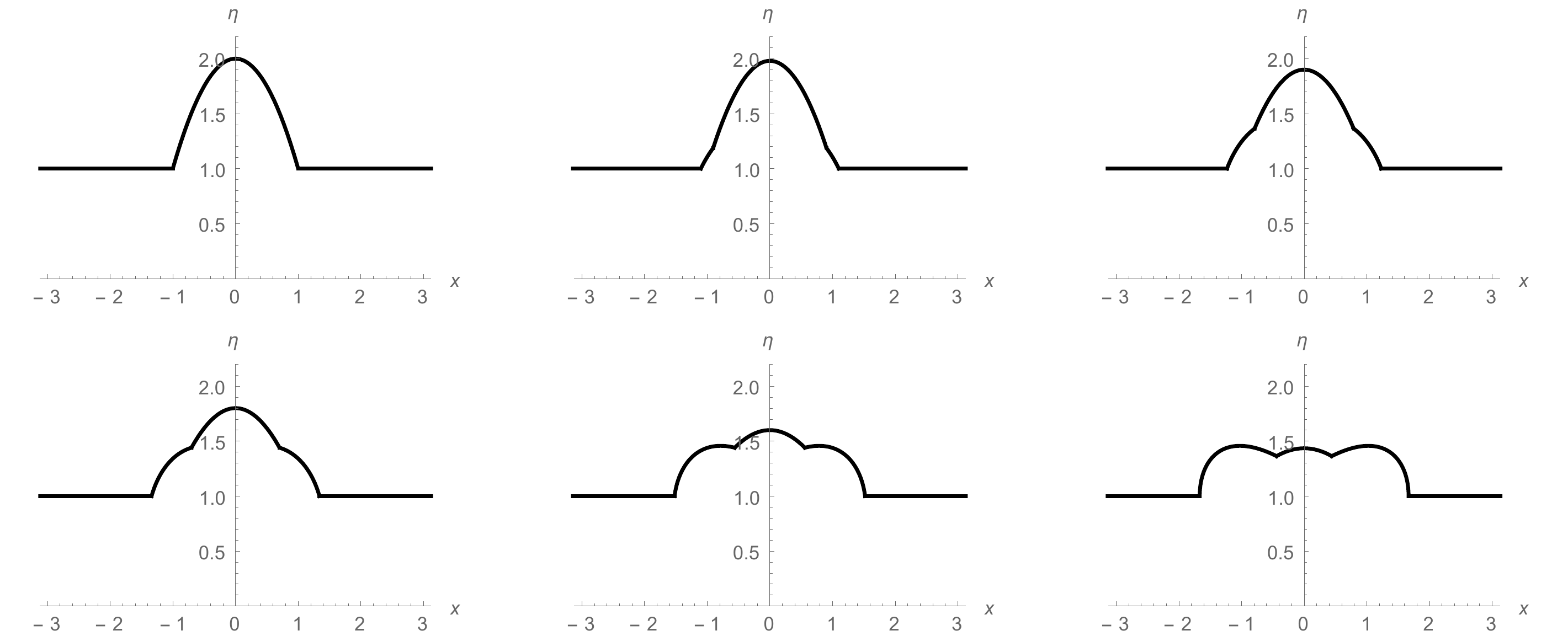}\\
        ({\it b}) \\
        \includegraphics[width=0.6\textwidth]{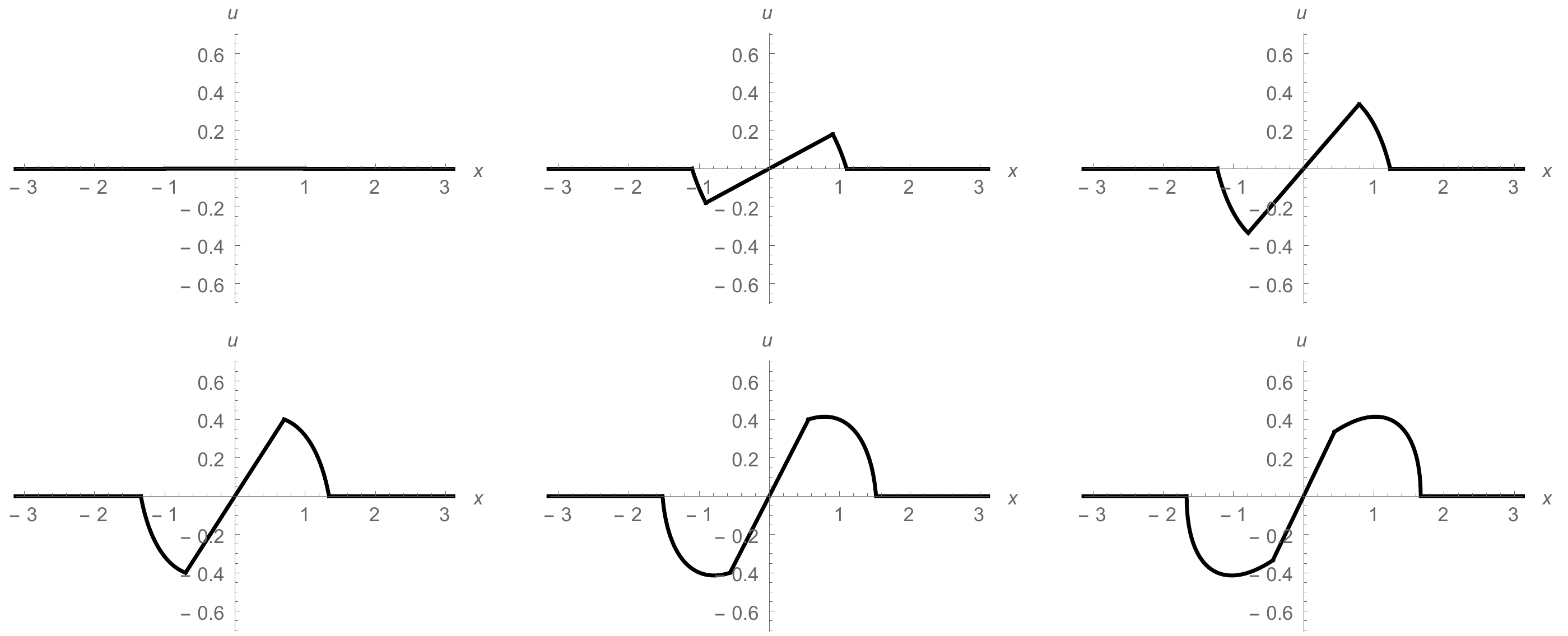}
    \caption{Time evolution of the water surface ({\it a}) and velocity field ({\it b}) stemming from initial conditions (\ref{pw initial eta}) and null initial velocity with a subcritical  value of the ratio $Q/\mu_0$.  Parameters are $Q=1$, $\gamma_0=-1$, $\mu_0=2$ and the snapshots are at time $t=0,\;0.1,\;0.23,\;0.34,\;0.52,\;t_\textnormal{sh}\simeq 0.67$.
  }
    \label{para-wet-eta-neg-fig}
\end{figure}
By introducing the auxiliary variable $\sigma=(\gamma/\gamma_0)^{1/3}$ in place of the time $t$, the  evolution of the parabolic part can be parametrized by
\begin{equation}
\label{parametric solution}
    \alpha=2\sqrt{|\gamma_0|}\, \sigma\sqrt{1-\sigma},\qquad \gamma=\gamma_0\sigma^3,\qquad \mu=\mu_0\sigma,
\end{equation}
with $\sigma(t)$ solving
\begin{equation}
\label{tau flat bottom}
    \dot \sigma=-2\sqrt{|\gamma_0|}\,\sigma^2\sqrt{1-\sigma},\qquad \sigma(0)=1,
\end{equation}
which can be integrated to give
\begin{equation}
\label{t(tau)}
    t=\frac{\sqrt{1-\sigma}+\sigma \arctanh \sqrt{1-\sigma}}{2\sqrt{|\gamma_0|}\;\sigma},
\end{equation}
This shows that $\sigma$ ranges monotonically from $1$ to $0$ as $t$ goes from $0$ to $+\infty$. In terms of the new variable $\sigma$, equation (\ref{dX_ell/dt}) turns into
\begin{equation}
    \frac{\de X_\ell}{\de \sigma}=-\frac{X_\ell}{\sigma}+\frac{1}{2}\sqrt{\frac{\sigma^2 X_\ell^2+(\mu_0/\gamma_0)}{\sigma^4-\sigma^3}},\qquad X_\ell(1)=x_0,
\end{equation}
which can be explicitly integrated by separation of variables with the substitution $w=\sigma X_\ell(\sigma)$. The solution is 
\begin{equation}
    X_\ell(\sigma)=\frac{\sqrt{(\mu_0-Q)\sigma}-\sqrt{Q(1-\sigma)}}{\sqrt{|\gamma_0|}\;\sigma}.
\end{equation}
\begin{figure}
    \centering
    ({\it a})\\
    \includegraphics[width=0.6
    \textwidth]{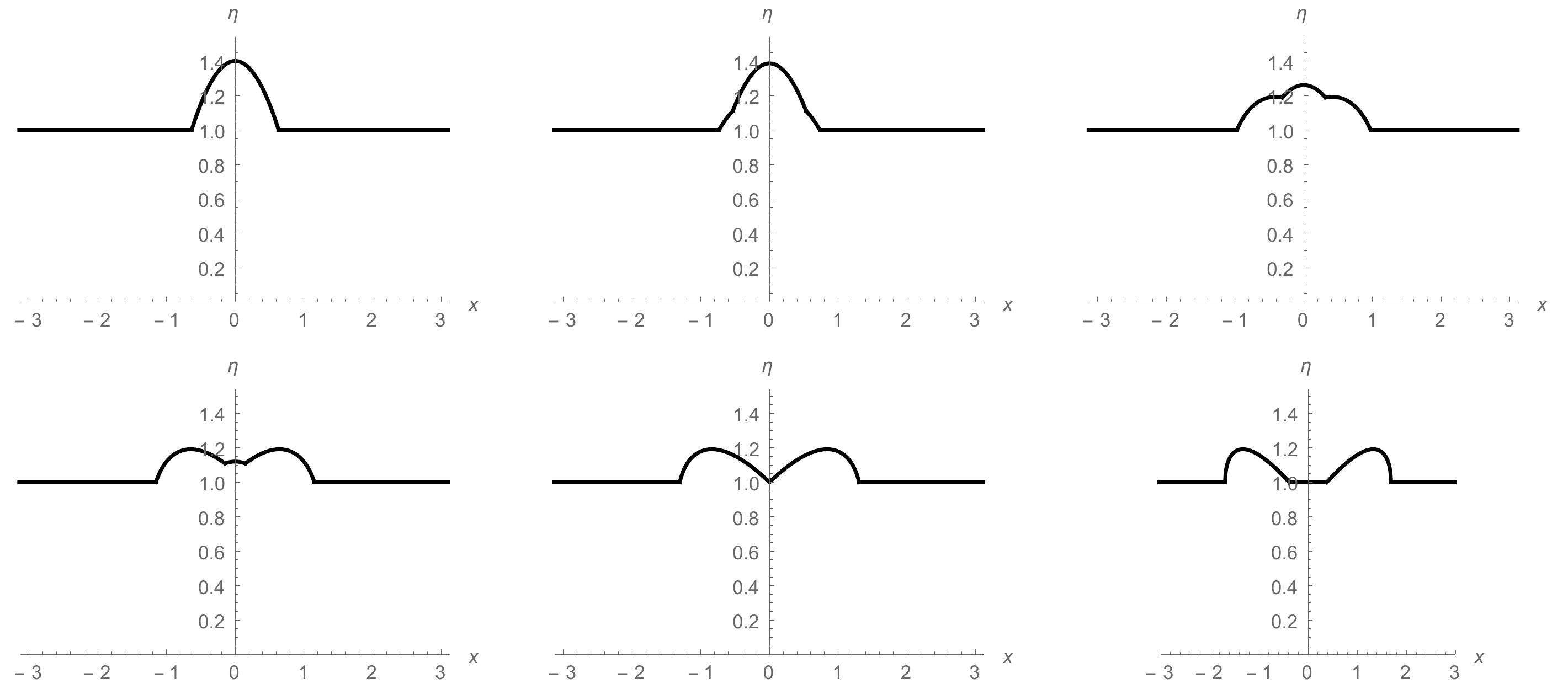}\\
    ({\it b})\\
        \includegraphics[width=0.6\textwidth]{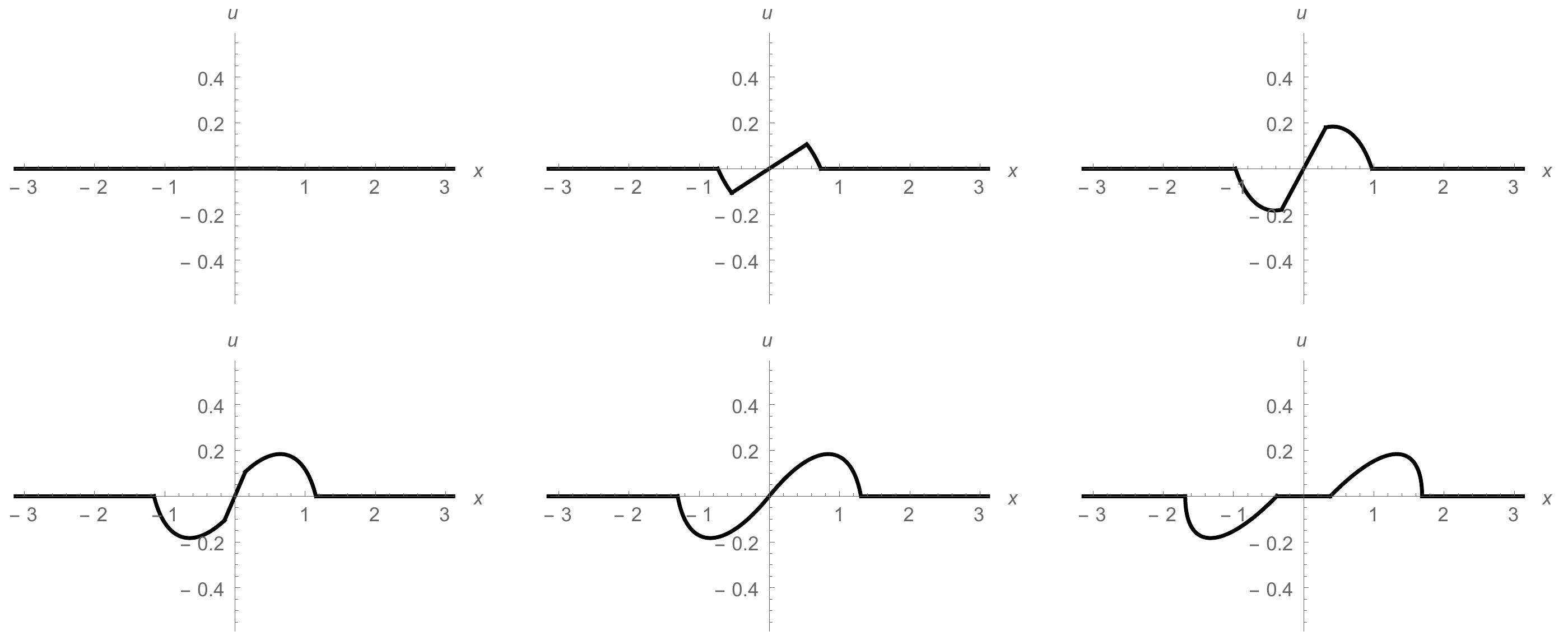}
    \caption{(a) Time evolution of the water surface (a) and velocity field (b) stemming from initial conditions (\ref{pw initial eta}) and null initial velocity with a supercritical value of the ratio $Q/\mu_0$. Parameters are $Q=1$, $\gamma_0=-1$, $\mu_0=1.4$ and the snapshots are at $t=0,\;0.1,\;0.34,\;0.52,\; t_c\simeq0.67,\;t_\textnormal{sh}\simeq 1.05$.}
    \label{para-wet-eta-neg-weak-fig}
\end{figure}
We are now in  position to apply the formalism of Section~\ref{sec: reminder flat bottom} and draw some conclusions about the shoulder part of the solution. This is given by relations (\ref{shoulder solution}),
\begin{equation}
\label{solution shoulder tau_0}
    N(\sigma_0)=\gamma(\sigma_0)X_\ell(\sigma_0)^2+\mu(\sigma_0),\qquad V(\sigma_0)=\alpha(\sigma_0)X_\ell(\sigma_0),
\end{equation}
with $\sigma(t_0)=\sigma_0$. As the left wavefront $x=X_\ell(t)$ propagates across the parabolic part of the solution, it eventually reaches the origin where it intersects the specular characteristic $x=-X_\ell(t)$ which originated at $x=-x_0$. We refer to the instant when this happens as the coalescence time. This instant is computed by setting $X_\ell(\sigma)=0$, which results in
\begin{equation}
    \sigma_c=Q/\mu_0.
\end{equation}
After setting $\sigma_0=\sigma_c$ in (\ref{solution shoulder tau_0}), we obtain
\begin{equation}
    N(\sigma_c)=Q, \qquad V(\sigma_c)=0,
\end{equation}
so that, at the coalescence time $t_c=(\sigma_c)$, both fields $\eta$ and $u$ regain their hydrostatic values at $x=0$, i.e., 
\begin{equation}
    \eta(0,t_c)=Q,\qquad u(0,t_c)=0.
\end{equation}
Therefore,  the parabolic part of the solution disappears at this time, and the background stationary state $\eta=Q$ and $u=0$ is attained at $x=0$. The solution features 
three singular points at $t=t_c$, since two of the previously existing ones, namely $x=\pm X_\ell(t)$, have coalesced into a single one, $x=0$. Subsequently, this singular 
point splits once again into a couple of new singular points which are transported along the characteristics passing through it. As hydrostatic conditions are established at 
$(0,t_c)$, the characteristics passing through this point are determined by
\begin{equation}
\label{new X_ell(t)}
    \frac{\de x}{\de t}=\pm\sqrt{Q}.
\end{equation}
Focusing once again on the positive part of the $x$ axis, we redefine the left boundary of the shoulder region for $t>t_c$ as the positive characteristic (\ref{new X_ell(t)}), defined  by
\begin{equation}
    x=X_\ell(t)=\sqrt{Q}\: (t-t_c)\qquad \textnormal{for}\; t>t_c.
\end{equation}
Note that, in terms of the parameter $\sigma_0$ (or $t_0$), this specific characteristic curve is identified by the label $\sigma_0=\sigma_c$ (or $t_0=t_c$).

The shock time can be explicitly calculated in this example as well. It is given by formulas (\ref{shock time single}) and (\ref{shock time inf}), expressed in terms of $\sigma$. Thus (\ref{shock time single}) reads
\begin{equation}
    \bar{\tau}(\sigma_0)=t_0+\frac{3\sqrt{N(\sigma_0)}-2\sqrt{Q}-\displaystyle 
   \frac{\de X_\ell}{\de \sigma}(\sigma_0)\frac{\de \sigma}{\de t}(t_0)}{3 \displaystyle \frac{\de \sqrt{N}}{\de\sigma}(\sigma_0)\frac{\de\sigma}{\de t}(t_0)},
\end{equation}
where $t_0=t(\sigma_0)$. The positive lower bound of these times, for $\sigma_c\le \sigma_0\le 1$, is  at $\sigma_0=1$, and evaluates to
\begin{equation}
   t_{\rm{sh}}= \tau_\textnormal{sh}=\frac{2}{3}\sqrt{\frac{Q/\mu_0}{|\gamma_0| (1-Q/\mu_0)}}.
   \label{tsh_expr}
\end{equation}
Therefore, a gradient catastrophe always happens for the class of initial conditions (\ref{pw initial eta}). Moreover, it takes place first on wavefront $x=X_r(t)$, as this is exactly the characteristic labelled by $\sigma_0=1$. Interestingly, the shock time can be either smaller or larger than the coalescence time,
\begin{equation}
    t_c=t(\sigma_c)=\frac{\displaystyle\sqrt{ 1-\frac{Q}{\mu_0}}+\displaystyle\frac{
    Q}{\mu_0}\arctanh\sqrt{1-\frac{Q}{\mu_0}}}{\displaystyle2\sqrt{|\gamma_0|}\frac{Q}{\mu_0}}.
    \label{tc}
\end{equation}
In fact, numerically computing the root $Q/\mu_0=\varrho\simeq 0.6213$ from the equality of $t_\textnormal{sh}$ from~(\ref{tsh_expr}) with $t_c$ from~(\ref{tc}), yields
\begin{equation}
    t_\textnormal{sh}\lesseqgtr t_c \iff \frac{Q}{\mu_0}\lesseqgtr \varrho\,,
\end{equation}
(with ${Q}/{\mu_0}\in(0,1)$).
In other words, if the initial parabolic core is ``steep" enough (i.e., the ratio $Q/\mu_0$ is small, not exceeding $\varrho$), the shock happens before the coalescence of the two singular points $x=\pm X_\ell(t)$, i.e. while the parabolic core of the solution is still present. This situation is depicted in Figures \ref{para-wet-eta-neg-fig}. In the opposite case, if the initial parabola is ``shallow," i.e.,  $Q/\mu_0>\varrho$, there is sufficient time  for the parabola to disappear completely at the coalescence time. Thereafter, the two symmetric shoulders move away from the origin in opposite directions until the shock develops. This situation is illustrated in Figures \ref{para-wet-eta-neg-weak-fig}.

\end{document}